\documentclass[useAMS,usenatbib]{mn2e}
\usepackage{rotating,times,pictex,graphicx,latexsym,color,pifont}

\title[Far-infrared photometry of deeply embedded outflow sources]{Far-infrared
photometry of deeply embedded outflow sources\thanks{Based on observations with
ISO, an ESA project with instruments funded by ESA Member States (especially
the PI countries: France, Germany, the Netherlands and the United Kingdom) and
with the participation of ISAS and NASA.}}

\author[D.~Froebrich, M.D.~Smith, K.-W.~Hodapp, and
        J.~Eisl\"offel]{D.~Froebrich,$^{1,4}$ M.D.~Smith,$^2$
        K.-W.~Hodapp$^3$, and
        J.~Eisl\"offel$^1$\\
        $^1$Th\"uringer Landessternwarte Tautenburg, Sternwarte 5, 
            D-07778 Tautenburg, Germany \\
        $^2$Armagh Observatory, College Hill, Armagh BT61 9DG,
            Northern Ireland\\
        $^3$University of Hawaii, Institute for Astronomy, 640 N. Aohoku Place,
            Hilo, HI 96720, USA\\
        $^4$Dublin Institute for Advanced Studies, 5 Merrion Square, Dublin 2,
            Ireland
        }
\date{Released 2003 Xxxxx XX}

\pagerange{\pageref{firstpage}--\pageref{lastpage}} \pubyear{2003}

\begin{document}

\label{firstpage} 

\maketitle

\begin{abstract} We present far-infrared maps and spectroscopy for a number of
deeply embedded protostellar objects (Cep\,E, HH\,211-MM, IC\,1396\,W, L\,1157,
L\,1211, and RNO\,15\,FIR) from data that we acquired with the ISO instruments
PHOT and LWS. Several previously undetected deeply embedded sources are found
in the vicinity of our  targets. We determine temperatures and luminosities of
seven objects and locate them on a L$_{bol}$-T$_{bol}$ diagram -- the
equivalent to a Hertzsprung-Russell diagram for protostars. Their masses and
ages, according to their location on tracks taken from our evolutionary model,
are derived. L\,1211 and Cep\,E appear to be intermediate mass objects which
will reach final masses of about 3\,M$_\odot$, while the other sources are in
or below the solar mass range. The derived ages of 15000 to 30000\,yr are
consistent with their current Class\,0 state. A comparison of the luminosity of
the associated outflows in the 1\,--\,0\,S(1) line of molecular hydrogen with
the source properties (bolometric luminosity, bolometric temperature, and
envelope mass) of 16 Class\,0 sources shows no statistically significant
correlations. Nevertheless, the data are consistent with a scheme in which the
outflow strength and protostar evolve simultaneously. We show that the
relationship is partially disguised, however, by the local properties of the
surrounding material, the extinction, and short-term flux variability.
\end{abstract}


%%%%%%%%%%%%%%%%%%%%%%%%%%%%%%%%%%%%%%%%%%%%%%%%%%%%%%%%%%%%%%%%%%%%%%%%%%%%%%
%%%%%%%%%%%%%%%%%%%%%%   Introduction  %%%%%%%%%%%%%%%%%%%%%%%%%%%%%%%%%%%%%%%
%%%%%%%%%%%%%%%%%%%%%%%%%%%%%%%%%%%%%%%%%%%%%%%%%%%%%%%%%%%%%%%%%%%%%%%%%%%%%%

\section{Introduction}

In the earliest stages of star formation -- the so-called Class\,0 and Class\,1
phases -- protostars are still deeply embedded in their parental molecular
cloud cores. This material absorbs almost all of the emitted radiation of the
star in the optical and the near-infrared. The spectral energy distribution
(SED) of Class\,0 protostars peaks at about 100--160\,$\mu$m, the wavelength of
the maximum of a modified blackbody at 30--80\,K. Hence, direct observations of
protostars have to be carried out in the far-infrared and in the (sub-)mm
wavelength range. Sub-mm and millimeter observations of some of the sources
investigated here have been obtained e.g. by Lefloch et al. \cite{lel96}, Ladd
\& Hodapp \cite{lh97}, Chini et al. \cite{cwknrs01}, Gueth et al.
\cite{ggdb97}, Motte \& Andr\'e \cite{ma01}, and Gueth \& Guilloteau
\cite{gg99}.

The ISO satellite (Kessler et al. \cite{ksa_etal96}) with its PHOT instrument
had the capacity to measure the broad-band continuum in the far-infrared. Such
observations, covering the peak region of the SED of protostars, help to yield
some of the major properties of these objects such as their temperature, the
sub-mm slope of their SED, the optical depth and the solid angle under which
they emit. The latter two cannot be disentangled due to the limited spatial
resolution of the ISOPHOT instrument. With higher resolution observations (e.g.
SCUBA), however, we can independently determine the solid angle under which an
object is seen and that way infer its optical depth. These parameters, together
with the distance, enable us to calculate the total (L$_{bol}$) and sub-mm
(L$_{smm}$) luminosities of each object. We may then decide whether an object
really is of Class\,0 or not by determining the L$_{smm}$/L$_{bol}$ ratio
(Andr\'e et al. \cite{awb00}). Finally, by placing the inferred values on a
temperature -- bolometric luminosity diagram -- the equivalent to a
Hertzsprung-Russell diagram for protostars (Myers et al. \cite{macs98}) -- we
are able for the first time to estimate the (model dependent) ages and masses
of these sources directly.

Bipolar outflows invariably accompany Class\,0 sources: strong inflow and
outflow of material are concurrent.  We thus wish to probe how the mass outflow
rate  is related to the mass accretion rate onto the protostar. The outflowing
material interacts with the ambient medium through  radiative shocks. Thus, the
luminosity of the outflow may be correlated with some of the source properties
(e.g. the bolometric source luminosity), which depend on the mass accretion
rate. Therefore, we measured the luminosities of the outflows of 16 Class\,0
sources in the 1\,--\,0\,S(1) line of molecular hydrogen. This is usually the
strongest  and easiest line to observe in near-infrared spectra of shocked
molecular hydrogen, and due to the short cooling time of H$_2$ it is a good
tracer of the present interaction of the outflow with the surrounding material.
These H$_2$ luminosities are then compared with various source properties to
investigate possible correlations.

Modelling of Class\,0 protostars remains in its infancy. Schemes now exist
which yield evolutionary tracks, based on relating gas accretion to the dusty
envelope (Myers et al.\cite{macs98}) and jet thrust to gas accretion (Bontemps
et al. \cite{batc96}, Saraceno et al. \cite{saraceno96}, Smith \cite{s98,s00},
and Andr\'e et al. \cite{awb00}). We combine these schemes here in order to
test if the simplest assumptions, such as a spherical envelope and a single
accreting object, are feasible.

In this paper, we first present our far-infrared ISO maps and spectroscopy, and
then summarize the data analysis and how we derive temperatures and
luminosities (Sect.\,\ref{datareduction}). In Sect.\,\ref{results}, we present
our results, and comment on individual objects. A discussion of age and mass
determination, and the general relationship to the outflows is contained in 
Sect.\,\ref{discussion}. A framework within which the data can be interpreted
is then put forward (Sect.\,\ref{evolscheme}).

%%%%%%%%%%%%%%%%%%%%%%%%%%%%%%%%%%%%%%%%%%%%%%%%%%%%%%%%%%%%%%%%%%%%%%%%%%%%%%
%%%%%%%%%%%%%  Observations and Data Reduction  %%%%%%%%%%%%%%%%%%%%%%%%%%%%%%
%%%%%%%%%%%%%%%%%%%%%%%%%%%%%%%%%%%%%%%%%%%%%%%%%%%%%%%%%%%%%%%%%%%%%%%%%%%%%%

\section{Observations and Data Analysis}

\label{datareduction}

We used the ISO satellite to obtain ISOPHOT minimaps of six Class\,0 sources
and LWS full grating spectra for three of them. All observations are listed in
Table\,\ref{obslog}.

\begin{table}
\begin{center}
\caption{\label{obslog} Log of our ISOPHOT and LWS observations.}
{\scriptsize
\begin{tabular}{clcccc}
Observation & Object & $\alpha$\,(J2000) &$ \delta$\,(J2000) & AOT &
t$_{exp}$\,[s] \\ 
number & & & & & \\
\noalign{\smallskip}
\hline
\noalign{\smallskip}
65903003&RNO\,15\,FIR&03 27 39&+30 13 00&PHT22& 670   \\
65903004&RNO\,15\,FIR&03 27 39&+30 13 00&PHT22& 620   \\
65903101&HH\,211     &03 43 57&+32 00 49&PHT22& 670   \\
65903102&HH\,211     &03 43 57&+32 00 49&PHT22& 620   \\
65201107&HH\,211     &03 43 57&+32 00 52&LWS01&2268~~~\\
65902801&HH\,211 West&03 43 57&+32 01 04&LWS01&3350~~~\\
66600502&HH\,211 East&03 43 59&+32 00 36&LWS01&2912~~~\\
46601429&L\,1157     &20 39 06&+68 02 13&LWS01&3390~~~\\
28200120&L\,1157     &20 39 06&+68 02 14&LWS01&1958~~~\\
52902105&L\,1157     &20 39 06&+68 02 14&PHT22& 668   \\
52902106&L\,1157     &20 39 06&+68 02 14&PHT22& 620   \\
54301407&IC\,1396W   &21 26 06&+57 56 17&PHT22& 668   \\
54301408&IC\,1396W   &21 26 06&+57 56 17&PHT22& 620   \\
56300709&L\,1211     &22 47 17&+62 01 58&PHT22& 670   \\
56300710&L\,1211     &22 47 17&+62 01 58&PHT22& 620   \\
56600912&Cep\,E South&23 03 13&+61 41 56&LWS01&1888~~~\\
56402111&Cep\,E      &23 03 13&+61 42 27&PHT22& 670   \\
56402112&Cep\,E      &23 03 13&+61 42 27&PHT22& 620   \\
56601113&Cep\,E North&23 03 13&+61 42 59&LWS01&1890~~~\\
\noalign{\smallskip}
\hline
\noalign{\smallskip}
\end{tabular}}
\label{tab1}
\end{center}
\end{table}

\subsection{ISOPHOT data}

Minimaps were taken for six objects (Cep\,E, HH\,211-MM, IC\,1396\,W, L\,1157,
L\,1211, and RNO\,15\,FIR) with ISOPHOT in its PHT22 mode by single pointing
and moving of the telescope by one (C100) or half (C200) of a detector pixel.
We used four filters (60, 100, 160, and 200\,$\mu$m). For 60 and 100\,$\mu$m,
the C100 detector (3$\times$3 array of Ge:Ga) was used to create a 5$\times$3
pixel minimap with a pixel size of 45\arcsec $\times$ 46\arcsec. The maps thus
cover a field of view of 230\arcsec $\times$ 135\arcsec. For the two longer
wavelengths 7$\times$3 mosaics with a pixel size of 45\arcsec $\times$
90\arcsec\, were obtained using the C200 detector (2$\times$2 array of stressed
Ge:Ga), covering thus a field of view of 315\arcsec $\times$ 270\arcsec. For
details on the instrument and the used Astronomical Observing Templates (AOT)
see the ISO Handbook, Volume V: PHT\,---\,The Imaging Photo
Polarimeter\footnote{http://www.iso.vilspa.esa.es/manuals/HANDBOOK/V/pht\_hb/}
and Lemke et al. \cite{lka_etal96}. The data were reduced with the ISOPHOT
Interactive Analysis (PIA V9.1) software. 

Flux measurements in the ISOPHOT maps were carried out in two different ways:
1) Point spread function (PSF) photometry using PSF fractions provided by
Laureijs \cite{l99} was done for the C100 maps. We do not provide PSF
photometry for the C200 detector since the given PSF fractions by Laureijs
\cite{l99} are only for the whole C200 pixel and our maps have a sampling of
half a pixel in one direction. 2) "Aperture" photometry was obtained for all
filters of both C100 and C200 detectors. Here we attributed each pixel in the
maps either to 'object' or to 'background' manually, then summed up both and
subtracted 'background' from 'object' to obtain its flux. Since at 60 and
100\,$\mu$m, i.e. for the C100 data, we were able to do photometry with both
methods, we have a means of estimating the consistency of both. All measured
fluxes, including the available IRAS fluxes of our objects, and the background
level in the maps are provided in Table\,\ref{phot_fluxes}.

\begin{table}
\renewcommand{\tabcolsep}{3pt}
\caption{\label{obslognir} Observation log of the NIR observations. The used
telescopes, detectors and filters are listed. H$_2$ indicates the narrow band
filter, centred at the 1\,--\,0\,S(1) line of H$_2$. The narrow band filter at
a wavelength of 2.140\,$\mu$m (continuum) is labeled with 2140. The number of
images is separately indicated for each filter. In some cases the
investigated objects fill only a part of the whole obtained mosaic (esp.
HH\,212). The observing time is given per single image.}
\centering
{\scriptsize 
\begin{tabular}{llllcc}
Observatory & Telescope & Object & Filter & Number of & t$_{obs}$(s) \\
Date & Detector &  &  & images &  \\
\noalign{\smallskip}
\hline
\noalign{\smallskip}
La Silla   & ESO/MPI\,2.2-m & HH\,24         & H$_2$, K'   & 12, 3~~    & 20, 2 \\
Apr93      & IRAC2          & Ser\,--\,FIRS1 & H$_2$, K'   & 23, 4~~    & 20, 2 \\
           &                & VLA\,1623      & H$_2$, K'   & 24, 4~~    & 20, 2 \\ \hline
Calar Alto & 2.2\,m         & L\,1448        & H$_2$, K'   & 32, 33     & 60, 3 \\
Jan94      & MAGIC          &                &             &            & \\ \hline
Calar Alto & 2.2\,m         & DR\,21         & H$_2$, K'   & 437, 78~~  & ~~25, 25 \\
Sep94      & MAGIC          & L\,1157        & H$_2$, K'   & 140, 13~~  & ~~~~25, 100 \\ \hline
Calar Alto & 3.5\,m         & Cep\,E         & H$_2$, 2140 & ~~52, 120  & ~~30, 30 \\
Nov95      & MAGIC          & HH\,211-MM     & H$_2$, 2140 & 60, 59     & ~~30, 30 \\
           &                & L\,1157        & H$_2$, 2140 & 44, 71     & ~~30, 30 \\ \hline
Mauna Kea  & UH\,2.2-m      & L\,1211        & H$_2$, K'   & 28, 27     & 200, 60 \\
Aug97      & QUIRC          &                &             &            &  \\ \hline
Calar Alto & 3.5\,m         & Cep\,A         & H$_2$, 2140 & 202, 368   & ~~20, 20 \\
Sep97      & MAGIC          & HH\,211        & H$_2$, 2140 & 32, 27     & ~~20, 20 \\
           &                & L\,1448        & H$_2$, K'   & 247, 253   & 20, 3 \\ \hline
Calar Alto & 1.2\,m         & Cep\,A         & H$_2$, K'   & 64, 32     & ~~60, 15 \\
Nov98      & MAGIC          & Cep\,E         & H$_2$, K'   & 34, 14     & ~~60, 15 \\
           &                & HH\,212        & H$_2$, K'   & 3112, 1487 & ~~60, 15 \\
           &                & L\,1448        & H$_2$, K'   & 400, 149   & ~~60, 15 \\
           &                & RNO\,15\,FIR   & H$_2$, K'   & 120, 60~~  & ~~60, 15 \\ \hline
Calar Alto & 3.5\,m         & Cep\,E         & H$_2$       & 24~~~~~~   & 30~~~~ \\
Dec00      & OMEGA PRIME    & HH\,211        & H$_2$       & 41~~~~~~   & 30~~~~ \\
           &                & L\,1157        & H$_2$       & 17~~~~~~   & 30~~~~ \\
           &                & L\,1448        & H$_2$       & 42~~~~~~   & 30~~~~ \\
\noalign{\smallskip}
\hline
\noalign{\smallskip}
\end{tabular}}
\end{table}

\subsection{LWS data}

For three objects (L\,1157, Cep\,E, and HH\,211) we have full grating
medium-resolution LWS01 scans, which cover a wavelength range from 43 to
196.9\,$\mu$m with a resolving power between 150 and 300. See the ISO Handbook,
Volume IV: LWS\,---\,The Long Wavelength Spectrometer\footnote{
http://www.iso.vilspa.esa.es/manuals/HANDBOOK/IV/lws\_hb/} and Clegg et al.
\cite{caa_etal96} for instruments and AOT details. We reduced the LWS data
using standard pipeline 7. For deglitching and flux calibration and defringing
of the spectra we employed the ISO Spectral Analyses Package (ISAP 1.6a). 

\subsection{Near-infrared H$_2$ observations}

For the measurement of the luminosities of the outflows in the 1\,--\,0\,S(1)
line of molecular hydrogen at 2.122\,$\mu$m near-infrared images were taken in
several observing campaigns and at various telescopes. The complete list of all
observations is provided in Table\,\ref{obslognir}. We observed the objects in
two filters to distinguish between line and continuum emission. Due to the
angular size of the objects, the single images had to be arranged into large
mosaics. All observing campaigns were (re)-reduced for consistency using 
own software based on the IRAF package DIMSUM. The whole procedure includes
flatfielding, cosmic ray hit removal and sky subtraction as well as
re-centering and mosaicing. For a higher astrometric accuracy we used all
available stars in the field for the re-centering. The photometric calibration
was achieved by the observation of faint near-infrared standards with an
accuracy of 10\%. For the flux measurements we subtracted the scaled continuum
image from the emission line image to measure only the flux in the
1\,--\,0\,S(1) line of H$_2$. 

Our images are being prepared for publication or are already  published.
Since in this paper we will only use the integrated H$_2$ line luminosities we
do not reproduce the images here. The objects are discussed in the following
papers: RNO\,15\,FIR in Davis et al. \cite{drec97} and Rengel et al.
\cite{rfhe02}; HH\,211-MM in  Eisl\"offel et al. \cite{efsm03}; VLA\,1623,
L\,1157 in Davis and Eisl\"offel \cite{de95}; L\,1211 in Froebrich and
Eisl\"offel \cite{fe04}; Cep\,E in  Eisl\"offel et al. \cite{esdr96}, Smith et
al. \cite{sfe03}; L\,1448 region in Eisl\"offel \cite{e00} and Froebrich et
al. \cite{fse02}; HH\,212, HH\,24 in Froebrich et al. \cite{fze01} and
Eisl\"offel et al. \cite{ezf04}; Ser-FIRS1 in Eisl\"offel and Froebrich
\cite{ef04}.

\subsection{Fit of the spectral energy distributions}
 
The observed broad-band continuum fluxes of our sources allow us to fit a SED
to the measurements and to infer source properties (e.g. T$_{bol}$ and
L$_{bol}$). To fit the SED we used Eq.\,\ref{graybody} for the flux density $S$
of our objects.
\begin{equation}
S[Jy]\,/\,\Sigma \Omega \, = \, \left(1-e^{-\tau}\right) \, \cdot \, B(\lambda,
T)
\label{graybody}
\end{equation}
$B(\lambda, T)$ is the Planck function, $\Sigma \Omega$ the solid angle  of the
source and $\tau$ the optical depth. $\tau$ is set as
\begin{equation}
\tau = \tau_{100} \cdot \left( \frac{\lambda}{100\mu m} \right) ^{-\beta}.
\end{equation}
$\lambda$ is in $\mu$m, the optical depth at 100\,$\mu$m ($\tau_{100}$) is a
free parameter, and $\beta$ is the sub-mm slope of the SED. The lowest $rms$ of
the fit is obtained when the solid angle of the object is determined by
\begin{equation}
\Sigma\Omega = \sum\limits_{f} \frac{(S_f / \Sigma\Omega) S_f^m}{(\Delta \,
S_f^m)^2} \left/ \sum\limits_{f} \frac{(S_f / \Sigma\Omega)^2}{(\Delta \,
S_f^m)^2} \right..
\label{sigmaomega}
\end{equation}
$f$ indicates the various used filters, $S^m_f$ the flux measurements in
these filters, and $\Delta \, S_f^m$ the error of the measurements. $S_f /
\Sigma\Omega$ is calculated by 
\begin{equation}
S_f / \Sigma\Omega = \frac{\int\limits_{\lambda = 0}^{\infty}
\left(1-e^{-\tau}\right) \, B(\lambda, T) \, T_f(\lambda) \,
d\lambda}{\int\limits_{\lambda = 0}^{\infty} T_f(\lambda) \, d\lambda} 
\label{convolve}
\end{equation}
for each filter $f$ separately using the filter transmission curves
$T_f(\lambda)$.

To fit a graybody to the measured SEDs of each object, a grid of graybodys was
computed (see Eq.\,\ref{graybody}) in which we varied the three parameters $T$,
$\tau_{100}$, and $\beta$. We varied the temperature between 15 and 80\,K, in
steps of 0.25\,K, the optical depth at 100\,$\mu$m from 0.09 to 40, in
logarithmic intervals of 1.5, and the sub-mm slope from 0.0 to 3.0, in steps of
0.1. These graybodys were convolved with the filter curves of the used filter
bands (see Eq.\,\ref{convolve}). Then the solid angle $\Sigma \Omega$ was
determined by computing the deviation of the model points from the measurements
and minimising this value (see Eq.\,\ref{sigmaomega}). Finally the $rms$ of the
fit to the measurements was calculated (see Eq.\,\ref{rms}; $n$ indicates the
number of filters) and the parameters leading to the minimal $rms$ were
selected. 

\begin{equation}
rms = \sqrt{ \frac{1}{n} \sum\limits_{f} \frac{(S_f(T, \beta, \tau_{100},
\Sigma\Omega) - S_f^m)^2}{(\Delta \, S_f^m)^2}} 
\label{rms}
\end{equation}

We find that $\tau_{100}$ has almost no influence on the shape of the graybody
curve, but only on the absolute flux level, which on the other hand mainly
depends on $\Sigma \Omega$. Thus, the values presented in
Table\,\ref{temperatures} are determined by fixing $\tau_{100}$ to unity. This
restriction has no influence in the deduced parameters T$_{bol}$ and L$_{bol}$,
but the given source size $\Sigma \Omega$ has no physical meaning. In
Sect.\,\ref{results} we present also graybody fits with $\tau_{100}$ as a free
parameter, in case that this improves the fit significantly. If sub-mm or
millimetre observations yield source sizes, the optical depth at 100\,$\mu$m
can be constrained.

Another way to determine physical meaningful radii for the sources is to
follow the assumptions of Myers et al. \cite{macs98}. They adopt an optically
thick envelope, a single power law dependence with the frequency of the
emissivity and an envelope density proportional to $r^{-3/2}$. Taking Figure\,2
in Myers et al. \cite{macs98}, we can determine the optical depth using $\log
\tau_{100} = - \log T_{bol}^2 + 4.28$. We used the bolometric temperatures from
Table\,\ref{temperatures} to derive $\tau_{100}$, and repeated the fit of the
SED with this optical depth. This leads to a new solid angle of the source
which is the size of the protostellar envelope where $\tau_{100}$ has the
correct value, according to the assumptions of Myers et al. \cite{macs98}. In
first approximation the solid angle and the optical depth are connected by
$\Sigma \Omega^{-1} \propto ( 1 - e^{- \tau_{100}})$. Thus, we can determine
the radius of the envelope where $\tau_{100}$ is unity (presented as R$_{100}$
in Table\,\ref{temperatures}). This radius is different from the envelope sizes
obtained of optically thin emission by sub-mm or millimeter measurements (e.g.
Motte \& Andr\'e \cite{ma01} and Chini et al. \cite{cwknrs01}).

\begin{figure}
\includegraphics[bb=140 325 375 550]{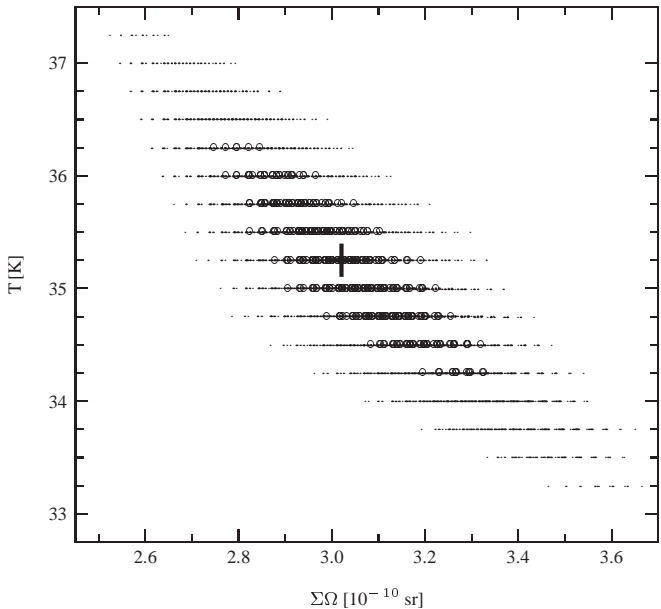}
\caption{\label{error} ($T$,$\Sigma \Omega$) plane for the graybody fit of the
Cep\,E photometry. The fit with the measured fluxes is marked by a cross.
Circles indicate fits using fluxes which deviate at most by 0.5\,$\sigma$, and
the small dots represent fits using fluxes with a maximum deviation of
1.0\,$\sigma$ from the measurements.}
\end{figure}

\begin{table*}
\renewcommand{\tabcolsep}{3pt}
\begin{center}
\caption{\label{phot_fluxes} Integrated far-infrared fluxes for all detected objects,
measured with ISOPHOT, and IRAS, as well as SCUBA and IRAM 30-m points from the
literature. Columns C$_{60}$..C$_{200}$ give the fluxes above the background
measured with aperture photometry, columns C$^{psf}_{60}$ and C$^{psf}_{100}$
the fluxes obtained by PSF fitting. For comparison we list the IRAS fluxes at
12, 25, 60, and 100\,$\mu$m in columns I$_{12}$..I$_{100}$. S$_{450}$ and
S$_{850}$ give SCUBA fluxes at 450 and 850\,$\mu$m from the literature.
I$_{1300}$ is the flux at 1.3\,mm. All fluxes are in Jansky. In the
B$_{60}$..B$_{200}$ columns we list the background level in the ISOPHOT maps at
60, 100, 160, and 200\,$\mu$m in MJy\,sr$^{-1}$. The * signs mark the
additional detected objects in our maps.} 
\begin{tabular}{llllllllllllllllll}
Object & C$_{60}$ & C$_{100}$ & C$_{160}$ & C$_{200}$ & C$_{60}^{psf}$ &
C$_{100}^{psf}$ & B$_{60}$ & B$_{100}$ & B$_{160}$ & B$_{200}$  & S$_{450}$ &
S$_{850}$ & I$_{1300}$ & I$_{12}$ & I$_{25}$ & I$_{60}$ & I$_{100}$ \\    
\noalign{\smallskip}
\hline
\noalign{\smallskip}
RNO\,15\,FIR            &25.7 &~~51.5 &~~46.7 &48.9 &24.8 &~~55.5 &~~19&~~~~25&~~~~92& ~~120& ~~9.2$^{(5)}$& ~~1.4$^{(5)}$&~~~~--         &0.25$^{(1)}$& 3.4&  47.1&~~93.6\\
RNO\,15$^{*}$           &~~3.3&~~11.9 &~~~~7.1&10.8 &~~3.7&~~10.5 &~~19&~~~~25&~~~~92& ~~120&~~~~--        &~~~~--        &~~~~--         &0.18$^{(1)}$& 4.2&  48.8&~~82.2\\
HH\,211-MM              &~~3.0&~~33.8 &~~56.2 &54.8 &~~1.9&~~20.8 &~~42& ~~138& ~~308& ~~331&  16.4$^{(5)}$& ~~3.8$^{(5)}$& ~~0.9$^{(2)}$ &~~--        &~~--&~~~~--&~~~~--\\
HH\,211\,FIRS2$^{*}$    &~~2.9&~~13.7 &~~55.3 &48.2 &~~1.0&~~11.3 &~~42& ~~138& ~~308& ~~331&~~~~--        &~~~~--        &~~~~--         &~~--        &~~--&~~~~--&~~~~--\\
L\,1157                 &~~6.8&~~37.8 &~~42.0 &38.6 &~~5.9&~~35.8 &~~11&~~~~19&~~~~63&~~~~72& ~~6.0$^{(6)}$& ~~0.9$^{(6)}$& ~~0.4$^{(6)}$ &0.25$^{(1)}$&0.25&  10.9&~~53.5\\
IC\,1396\,W             &~~5.7&~~19.9 &~~36.6 &26.6 &~~4.6&~~13.8 &~~25&~~~~92& ~~210& ~~230&~~~~--        &~~~~--        &~~~~--         &0.25$^{(1)}$& 0.6&~~9.7 &~~38.3\\
IC\,1396\,W\,FIRS2$^{*}$&~~0.1&~~~~1.2&~~33.2 &~~9.9&~~0.2&~~~~1.2&~~25&~~~~92& ~~210& ~~230&~~~~--        &~~~~--        &~~~~--         &~~--        &~~--&~~~~--&~~~~--\\
IC\,1396\,W\,FIRS3$^{*}$&~~0.1&~~~~1.2&~~~~4.6&17.3 &~~0.2&~~~~2.3&~~25&~~~~92& ~~210& ~~230&~~~~--        &~~~~--        &~~~~--         &~~--        &~~--&~~~~--&~~~~--\\
L\,1211                 &12.9 &~~36.5 &~~63.0 &75.0 &10.5 &~~23.4 &~~42& ~~157& ~~254& ~~280&~~~~--        &~~~~--        &~~0.135$^{(3)}$& 2.7& 5.7&  19.8&~~63.6$^{(1)}$\\
L\,1211\,FIRS2$^{*}$    &~~3.4&~~17.0 &~~31.6 &46.2 &~~2.2&~~10.9 &~~42& ~~157& ~~254& ~~280&~~~~--        &~~~~--        &~~0.345$^{(4)}$&~~--        &~~--&~~~~--&~~~~--\\
Cep\,E                  &55.2 &125.8  &102.1  &81.4 &65.4 &123.0  &~~35&~~~~99& ~~234& ~~293&  43.7$^{(6)}$& ~~4.1$^{(6)}$& ~~1.0$^{(6)}$ &0.43& 5.8&  61.0&112.0 \\
\noalign{\smallskip}
\hline
\end{tabular}
\begin{list}{}{}
\item[(1)] upper limit \,\,(2) measured at 1.1\,mm (McCaughrean et al.
\cite{mrz94}) \,\,(3) measured at 1.2\,mm (Tafalla et al. \cite{tmmb99}),
MMS\,4 \,\,(4) measured at 1.2\,mm (Tafalla et al. \cite{tmmb99}),
superposition of MMS\,1, MMS\,2 and MMS\,3 \,\,(5) Rengel et al. 
\cite{rfeh01} \,\,(6) Chini et al. \cite{cwknrs01}
\end{list}
\end{center}
\end{table*}
 
The above described method to fit the SED was applied to both, the measurements
at our four ISOPHOT wavelengths and all available data from
Table\,\ref{phot_fluxes} (except the IRAS 12 and 25\,$\mu$m points; see below).
The inferred object properties from the latter are listed in
Table\,\ref{temperatures}. We did not find significant changes in the fit
parameters between the two methods, except for L\,1211 and partly for
RNO\,15\,FIR. These differences are discussed in Sect.\,\ref{res_l1211} and
\ref{res_rno15fir}, respectively. All obtained results are discussed for each
object separately in Sect.\,\ref{results}.

IRAS 12 and 25\,$\mu$m points are excluded from the fit of the SED since recent
works show, that these fluxes are usually far above the fit to the SED (see
e.g. Chini et al. \cite{cwknrs01}, Barsony et al. \cite{bwao98}). Barsony et
al. \cite{bwao98} argue that this excess mid-IR emission is due to ongoing
outflow/dust interactions. Another effect, leading to the detection of these
extremely red sources at such short wavelength (and especially at 12\,$\mu$m)
could be the tiniest of red leaks of the IRAS filters. Suppose that there was a
red leak of only 0.1\,\% in the 12\,$\mu$m filter -- at the limit of which the
filter transmission curve is known -- over the band pass between 20 and
30\,$\mu$m. The IRAS detectors were still sensitive in this range. Such leak
would increase the measured I$_{12}$ flux of Cep\,E, for example, by a factor
of 500. We also note that the 12 and 25\,$\mu$m IRAS data points, if used, for
most deeply embedded sources show huge deviations from the best-fitting SEDs,
compared to the presumed precision of the flux calibration of IRAS. Less red
sources, like Class\,1 or 2 objects, would hardly suffer from such leaks,
because their SEDs are a lot less steep in the 12 and 25\,$\mu$m range.
Therefore, we decided not to use the IRAS 12 and 25\,$\mu$m data points in our
SED fits for our very red objects.

If one uses the measurements for the determination of the bolometric
temperature including also the 12 and 25\,$\mu$m IRAS points, nevertheless, we
get slightly higher values (about 4\,K). Since it is not clear how much of this
small effect is due to filter leaks, outflow/dust interactions, or envelope
emission, we do not list these values in Table\,\ref{temperatures}.

Our best fitting results are given in Table\,\ref{temperatures} together with
the fit errors. In almost all cases it was not possible or not useful to do a
fit for the newly detected objects in our maps, since they are at the edge of
the map and so we are missing an unknown part of their flux. Some objects are
detected only with the C200 detector since they are outside the slightly
smaller maps at the C100 wavelengths.

A determination of the fit errors cannot be obtained analytically. Therefore,
we varied the measurements within their one sigma error box (five equidistant
values; $S_f^m \pm n/2 \cdot \Delta S_f^m$; n = 0, 1, 2) and computed the best
fitting parameters for each of the $5^7 = 78125$ combinations. This results in
an area of the parameter space into which the error boxes are mapped. As an
example, we show in Fig.\,\ref{error} this area in the ($T$,$\Sigma \Omega$)
plane for Cep\,E. The errors given in Table\,\ref{temperatures} are read off
such diagrams for each of our objects. This procedure was applied also to the
parameters L$_{bol}$, L$_{smm}$ and T$_{bol}$.

The determined graybody fits are integrated to obtain the total luminosities of
the sources. By integrating only at wavelengths larger than 350\,$\mu$m we
obtain the sub-mm luminosities L$_{smm}$, which can be compared to the total
luminosities L$_{bol}$ to decide whether an object is a Class\,0 source
(Andr{\'e} et al. \cite{awb00}). Both values, L$_{bol}$ and
L$_{smm}$/L$_{bol}$, are given in Table\,\ref{temperatures}. When the ratio
L$_{smm}$/L$_{bol}$ exceeds 0.005, then the object is counted as Class\,0. This
is equivalent to the mass ratio M$_{env}$/M$_*$ being larger than unity (see
Andr{\'e} et al. \cite{awb00} and references therein). The given bolometric
temperatures T$_{bol}$ are the temperatures of a blackbody with the same mean
frequency as the graybody, where the mean frequency $\overline\nu$ of an SED
is determined by

\begin{equation}
\overline\nu = \int\limits_{0}^{\infty} \nu \cdot \mbox{SED}(\nu)\,\, d\nu
\left/ \int\limits_{0}^{\infty} \mbox{SED}(\nu)\,\, d\nu \right.
\end{equation}
 
\begin{table*}
\begin{center}
\caption{\label{temperatures} Best graybody fit results, as well as the
inferred bolometric and sub-mm luminosities, from our ISOPHOT data and (if
available) SCUBA and millimeter measurements from the literature. T is the
fitted temperature of the graybody, $\beta$ the sub-mm slope of the SED, and
$\Sigma \Omega$ the solid angle of the source. The optical depth at 100\,$\mu$m
was fixed to 1.0, since it did not show significant influence on the shape of
the graybody. R$_{100}$ gives the corresponding radius of the envelope where
$\tau_{100}$ is unity, determined with T$_{bol}$ and following the assumptions
of Myers et al. (1998). A discussion of fits with variable $\tau_{100}$ can be
found in Sect.\,\ref{results}. The $rms$ gives the deviation of the fit from
the measurements scaled with the errors of the measurements (see
Eq.\,\ref{rms}). The explanation of the determination of the errors is given in
the text. $\Sigma \Omega$ is given in 1$\times$10$^{-10}$\,sr (equal to
4.25\,$\Box$\arcsec). The sub-mm luminosity L$_{smm}$ is the luminosity of the
object at wavelengths larger than 350\,$\mu$m, and the bolometric temperature
T$_{bol}$ is the temperature of a blackbody with the same mean frequency as the
object. The * sign marks a newly discovered object. Due to the photometry
problems with this object, we do not present errors here.}
\begin{tabular}{lccccccccccccc}
Object & T\,[K] & $\beta$ & $\Sigma \Omega$ & R$_{100}$\,[AU] & $rms$ & R\,[pc]
& T$_{bol}$\,[K] & L$_{bol}$\,[L$_{\odot}$] & L$_{smm}$/L$_{bol}$ & Class\,0\\   
\noalign{\smallskip}
\hline
\noalign{\smallskip}
RNO\,15\,FIR            &34.0$\pm$3.0 &1.1$\pm$0.3 &1.7$\pm$0.2 &270&1.0  &350 &44.6$\pm$3.0 &~~8.4$\pm$1.0 &0.017$\pm$0.007 &\ding{51} \\
HH\,211-MM              &21.0$\pm$3.0 &1.5$\pm$0.6 &12.5$\pm$1.0&520&2.0  &315 &31.4$\pm$1.0 &~~4.5$\pm$0.5 &0.046$\pm$0.020 &\ding{51} \\
L\,1157                 &26.5$\pm$1.5 &1.4$\pm$0.4 &3.3$\pm$0.3 &440&1.5  &440 &37.8$\pm$1.5 &~~7.6$\pm$0.8 &0.025$\pm$0.015 &\ding{51} \\
IC\,1396\,W             &30.0$\pm$2.0 &0.3         &1.1$\pm$0.2 &680&1.3  &750 &32.6$\pm$2.0 &16.4$\pm$2.0  &0.059$\pm$0.010 &?         \\
L\,1211                 &30.5$\pm$2.0 &0.0         &2.1$\pm$0.6 &280&1.4  &725 &30.5$\pm$2.0 &33.1$\pm$4.0  &0.073$\pm$0.012 &?         \\
L\,1211\,FIRS2$^{*}$    &26.8         &0.0         &1.7         &340&1.6  &725 &26.9         &16.0          &0.100           &?         \\
Cep\,E                  &35.3$\pm$3.0 &1.0$\pm$0.3 &3.0$\pm$0.4 &750&0.9  &730 &45.0$\pm$3.0 &77.9$\pm$10   &0.017$\pm$0.010 &\ding{51} \\
\noalign{\smallskip}
\hline
\noalign{\smallskip}
\end{tabular}
\end{center}
\end{table*}
 
%%%%%%%%%%%%%%%%%%%%%%%%%%%%%%%%%%%%%%%%%%%%%%%%%%%%%%%%%%%%%%%%%%%%%%%%%%%%%%
%%%%%%%%%%%%%%%%%%%%%%   Results    %%%%%%%%%%%%%%%%%%%%%%%%%%%%%%%%%%%%%%%%%%
%%%%%%%%%%%%%%%%%%%%%%%%%%%%%%%%%%%%%%%%%%%%%%%%%%%%%%%%%%%%%%%%%%%%%%%%%%%%%%

\section{Results}

\label{results}

Our observations of Cep\,E, HH\,211-MM, IC\,1396\,W, L\,1157, L\,1211, and
RNO\,15\,FIR were carried out at their nominal IRAS positions. In our ISOPHOT
maps (shown in Figs.\,\ref{cepe_data}\,--\,\ref{rno15fir_data}) we detected
more objects than were actually targeted. In four cases other (partly)
unexpected embedded objects or bright diffuse continuum emission are found. For
L\,1211 no object was detected at the nominal IRAS position, but there were two
other sources discovered in the maps. Measured fluxes in all filters, including
IRAS fluxes and sub-mm and millimeter points from the literature, are given in
Table\,\ref{phot_fluxes}.

Discrepancies of the fluxes between PSF and "aperture" photometry are for
various reasons: First, it is a major problem to determine which pixel
contributes to which object when doing "aperture" photometry. A second problem
is the determination of the background. When using the PSF fitting method, the
background is determined automatically (provided that the object is a point
source and in the centre of a pixel), while for "aperture" photometry one has
to choose background pixels. Concerning the absolute calibration errors for the
two detectors of 15 and 10\% for the C100 and C200 detector, respectively, and
an additional error of 20\% due to background uncertainties, we find that both
flux determination methods lead to consistent results in almost all cases.

Most of our investigated objects are of Class\,0 type according to the
L$_{smm}$/L$_{bol}$ criterion. We cannot decide whether the newly discovered
objects in our maps are of Class\,0, because they are situated at the edges of
the ISOPHOT maps. Due to the different sizes of the maps we certainly
underestimate their fluxes at 60 and 100\,$\mu$m, which alters their derived
SED in the way that they seem to be proportionally brighter at the longer
wavelengths, but to an unknown extent. 

The ISOPHOT and the IRAS fluxes at 60 and 100\,$\mu$m are consistent within the
errors only for Cep\,E. For all other objects the IRAS point source catalogue
gives values which are a factor of about 1.8 brighter. Apart from the fact that
the errors for the IRAS data are quite large and in some cases only upper
limits are given, the main reason for the differences is that the resolution of
the IRAS satellite was not sufficient to resolve close-by sources. Only Cep\,E
and L\,1157 seem not to have other young objects in their immediate vicinity,
and these are the two objects where the IRAS and ISOPHOT fluxes match the best.
Cep\,E is a known double source (Moro-Mart\'{\i}n et al. \cite{mnmtcs01}) which
cannot be resolved by IRAS nor ISOPHOT. For these reasons and the still fairly
large errors in the flux measurements, no investigation of the time evolution
of the fluxes of these young sources over the 14 year time span ($\approx$
0.1\% of the age of our objects) between IRAS and ISO is possible. 

The PSF photometry suggests that all the objects are seen as point sources for
the ISOPHOT detectors. When subtracting the fitted PSF, no systematic residuals
are visible in the difference images. Thus, the angular size of the sources is
at maximum 10\arcsec, a quarter of the FHWM of the PSF. This leads to an upper
limit for the source solid angles of about 100$\Box$\arcsec. This fact is
supported by the inferred sizes R$_{100}$ on the order of
1$\times$10$^{-10}$\,sr (4\,$\Box$\arcsec), which is less than one percent of
the pixel size of the C100 detector.

For the three objects for which we obtained a LWS spectrum, we can compare the
PHOT flux with the LWS continuum. While the LWS continuum is a sum of the
continuum of the source and background radiation, the PHOT maps give the true
flux of the source. So, the difference between LWS and PHOT should be the
background radiation (e.g. from cold dust). In all three cases (Cep\,E,
L\,1157, and HH\,211-MM) we clearly see evidence for such a background emission
(see Figs.\,\ref{cepe_data}\,--\,\ref{hh211_data}).

In the following subsections we discuss details of the results for the
individual objects.

\subsection{Cep\,E}

\begin{figure}
\includegraphics[bb=140 365 370 510]{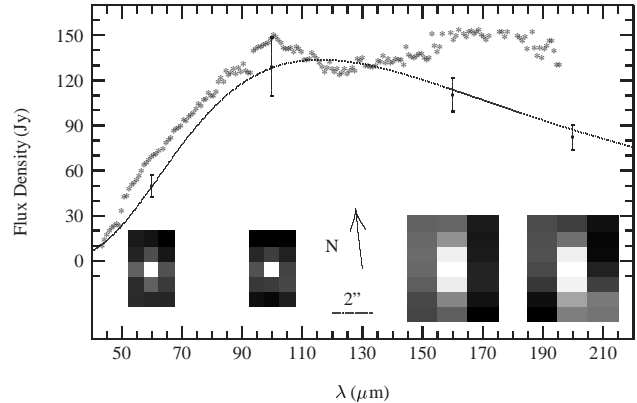}
\caption{\label{cepe_data} ISOPHOT maps and derived photometry for Cep\,E.
Stars (*) show the LWS spectrum and the solid line the best graybody fit
(T\,=\,35.3\,K, $\beta$\,=\,1.0, $\Sigma
\Omega$\,=\,3.0\,$\times$10$^{-10}$\,sr) to the data. $\tau_{100 }$ is fixed to
unity. Error bars for the PHOT data only include the 15\% and 10\% errors of
the detectors. They do not include additional uncertainties due to background
determination. The four maps are centred on the central wavelength of the
appropriate filter and have the same scale and orientation, given by the scale
and the arrow.}
\end{figure}

Cep\,E is the brightest object in our sample. Our ISOPHOT maps at the four
wavelengths of 60, 100, 160, and 200\,$\mu$m are shown in the lower part of
Fig.\,\ref{cepe_data}, all at the same scale and orientation. Photometry from
these maps and the LWS spectrum, are displayed above the maps. In addition, we
plot the best-fitting graybody to these data as solid line. The fit was used to
deconvolve the measurements and the filter transmission curve for converting
the measured fluxes to flux densities at the central wavelengths of the used
filter. For Cep\,E, the fluxes determined with PSF and "aperture" photometry
were consistent. Deviations of the LWS continuum from the ISOPHOT data exist
for wavelengths shorter than 100 and longer than 150\,$\mu$m. This might be
evidence for warm and cold dust. 

The PSF photometry shows that the object is a point source, perfectly aligned
in the middle of our map, and no other embedded object is detected.
Nevertheless Cep\,E is at least a double source, separated by 1.4\arcsec\, 
($\approx$\,1000\,AU), as shown by the 222\,GHz observations of
Moro-Mart\'{\i}n et al. \cite{mnmtcs01}.

Cep\,E was observed by Chini et al. \cite{cwknrs01} with SCUBA (450 and
850\,$\mu$m) and the IRAM 30-m telescope (1.3\,mm). They measured the fluxes in
an aperture with a radius of 40\arcsec, comparable to the size of our ISOPHOT
pixels. The fluxes are given in Table\,\ref{phot_fluxes}. We included these
data in the graybody fit. The deduced temperature is 35.3\,K for the best
graybody fit, which also gives $\beta$\,=\,1.0 and $\Sigma
\Omega$\,=\,3.0$\times$10$^{-10}$\,sr. These parameters are computed fixing
$\tau_{100}$ to unity. If we vary the optical depth also, the $rms$ of the fit
is lowered from 0.9 to 0.7. The new graybody parameters are then:
T\,=\,42.8\,K, $\beta$\,=\,1.5, $\Sigma \Omega$\,=\,0.8$\times$10$^{-10}$\,sr
and $\tau_{100}$\,=\,25. Nevertheless, the inferred bolometric luminosities and
bolometric temperatures are not affected. We obtain 77.9, 79.6\,L$_\odot$ and
45.0, 45.7\,K for a fixed and free $\tau_{100}$, respectively. The
L$_{smm}$/L$_{bol}$ ratio is 0.017, a strong hint for the Class\,0 nature of
this object, even if we observe the superposition of two sources. Using the
assumptions of Myers et al. \cite{macs98} and T$_{bol}$ we calculate an optical
depth at 100\,$\mu$m of 9.4. From this we determine a radius of the
protostellar envelope where $\tau_{100}$ is unity of 750\,AU. Thus, the
diameter (1500\,AU) is in agreement with Moro-Mart\'{\i}n et al.
\cite{mnmtcs01} who found that the double system (separation of 1000\,AU) is
surrounded by a common envelope. 

The bolometric temperature of about 45\,K is well below the value of 60\,K
given by Ladd and Hodapp \cite{lh97}. They used the IRAS data (12, 25, 60, and
100\,$\mu$m) and an 800\,$\mu$m point to fit the bolometric temperature. Chini
et al. \cite{cwknrs01} could fit the 100\,$\mu$m IRAS measurement and the 450,
850 and 1300\,$\mu$m points. These data still do not cover the emission maximum
of the source at about 130\,$\mu$m. For an accurate determination of the
temperature, however, the position of the maximum of the SED is needed, which
was observed here with ISOPHOT.

\subsection{L\,1157}

\begin{figure}
\includegraphics[bb=140 365 370 510]{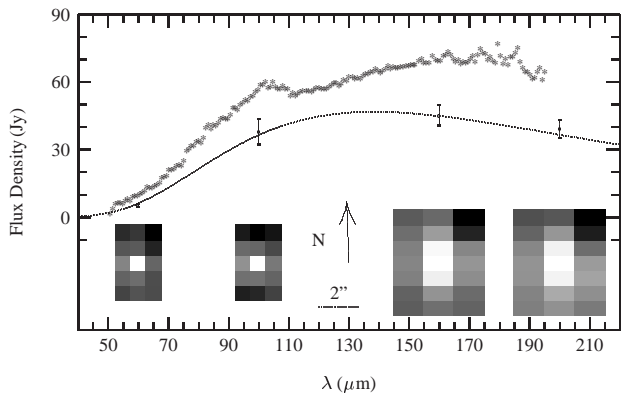}
\caption{\label{l1157_data} As Fig.\,\ref{cepe_data}, but for L\,1157. The
best graybody has the parameters T\,=\,26.5\,K, $\beta$\,=\,1.4, and $\Sigma
\Omega$\,=\,3.3$\times$10$^{-10}$\,sr. $\tau_{100}$ was fixed to unity.}
\end{figure}

\begin{figure*}
\centering
\parbox[h]{18cm}{
\includegraphics[bb=145 305 625 570]{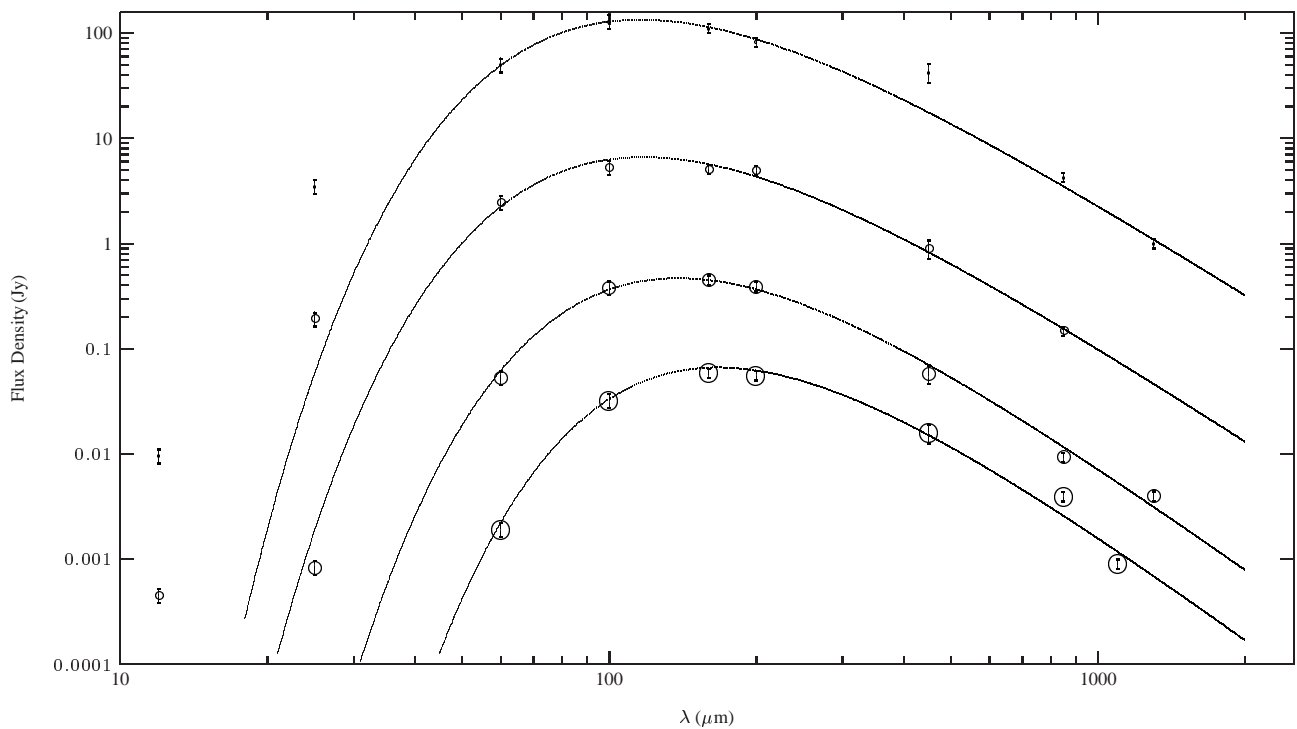}
}\bigskip\\
\caption{\label{scuba_data}
Best obtained fit to the SED using Eq.\,\ref{graybody}. Overplotted are the
IRAS (12 and 25\,$\mu$m), ISOPHOT (60, 100, 160, and 200\,$\mu$m), SCUBA (450
and 850\,$\mu$m), and millimeter (1100, 1300\,$\mu$m) data points of Cep\,E,
RNO\,15\,FIR, L\,1157, and HH\,211-MM (from top to bottom and from small to big
circles). $\tau_{100}$ was fixed to unity. For the obtained parameters of the
best fit see text or Table\,\ref{temperatures}. The models and datapoints are
shifted for RNO\,15\,FIR, L\,1157, and HH\,211-MM by one, two and three orders
of magnitude down, respectively, for convenience. HH\,211-MM was not detected
by IRAS.}
\end{figure*}
 
Our PHOT maps of L\,1157, and the integrated photometry obtained from these
maps, are presented in Fig.\,\ref{l1157_data}. This figure also shows our LWS
spectrum of L\,1157, as well as the graybody fit to the photometry.  L\,1157
was observed with SCUBA and IRAM 30-m by Chini et al. \cite{cwknrs01}. They
give two different measurements for the fluxes: one for the central source only
(10\arcsec\, aperture) and one for the source and the whole envelope (a
55\arcsec\, by 30\arcsec\, elliptical aperture). The fluxes for the central
source are given in Table\,\ref{phot_fluxes}.

L\,1157 is a point source with a derived temperature of about 26.5\,K,
$\beta$\,=\,1.4, and $\Sigma \Omega$\,=\,4.3$\times$10$^{-10}$\,sr, under the
assumption that $\tau_{100}$ is unity. Varying the optical depth does not
improve the fit. The bolometric temperature and luminosity are 37.8\,K and
7.6\,L$_\odot$, while 2.5\% of these account for L$_{smm}$. This classifies
L\,1157 as a Class\,0 source. According to the scheme of Myers et al.
\cite{macs98} we determine $\tau_{100} = 13$ and R$_{100} = 440$\,AU. If we use
the fluxes given by Chini et al. \cite{cwknrs01} for the envelope, we get a
much worse fit and it seems that we have a second cool component in the SED.
Thus, the ISOPHOT data reflect the emission of the source itself and not the
cold extended envelope.

There are small deviations of the LWS continuum from the PHOT photometry over
the whole wavelength range. This could be due to diffuse emission from warm and
cold dust, or reflecting uncertainties in the calibration of ISOPHOT or LWS.
PSF photometry shows that the object is not at the centre of our map, but
rather shifted slightly to the east. This could be a hint for another source
nearby or a slight mispointing of the telescope due to the limited accuracy of
the IRAS coordinates. Since nothing is known in the literature about a second
source, we attributed all the flux to L\,1157. PSF fitting to the C200 maps to
confirm this was not possible, due to the unknown PSF for our half-pixel
sampling in the north-south direction. Also the inferred size of the source
($\Sigma \Omega$) does not support the presence of an additional object.

\subsection{HH\,211-MM}

\begin{figure}
\includegraphics[bb=140 355 370 520]{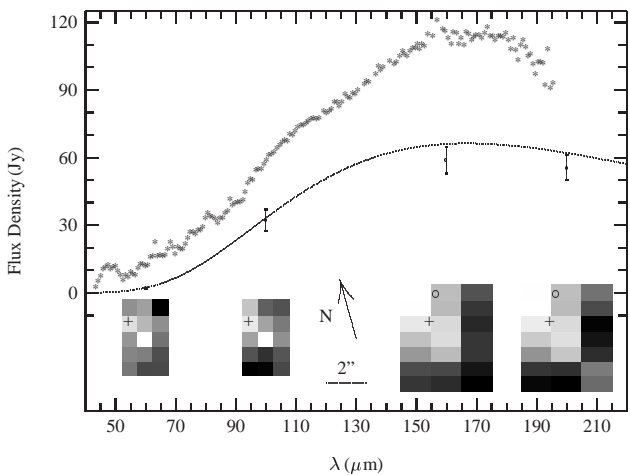}
\caption{\label{hh211_data} 
As Fig.\,\ref{cepe_data}, but for HH\,211-MM. The best graybody has the
parameters T\,=\,21.0\,K, $\beta$\,=\,1.5, and $\Sigma
\Omega$\,=\,12.5$\times$10$^{-10}$\,sr. $\tau_{100}$ was fixed to unity. The
position of IC\,348\,IR is indicated by a + sign, and a circle marks the source
IC\,348\,MMS, found by Eisl\"offel et al. (2003).}
\end{figure}

Our PHOT maps, derived photometry and a LWS spectrum of the HH\,211 region are
displayed in Fig.\,\ref{hh211_data}. HH\,211-MM at the centre of our maps is
the dominant source at 60 and 100\,$\mu$m. IC\,348\,IR, probably a heavily
embedded B-star (Strom et al. \cite{ssc74}, McCaughrean et al. \cite{mrz94}),
is visible to its north-east (marked by a cross). At longer wavelengths, a very
cold source HH\,211\,FIRS2 further north becomes visible and even dominant. It
probably coincides with the object IC\,348\,MMS (marked by a circle), found by
Eisl\"offel et al. \cite{efsm03} to be the source of a newly detected outflow
north of HH\,211. Thus, the fluxes of HH\,211\,FIRS2 given in
Table\,\ref{phot_fluxes} are a superposition of two different objects. The
C$_{60}$ and C$_{100}$ measurements are dominated by IC\,348\,IR, while
C$_{160}$ and C$_{200}$ are dominated by IC\,348\,MMS. Therefore no further
investigation of the SED of one of these objects was possible.

For HH\,211-MM, we find large differences of the fluxes at 60 and 100\,$\mu$m
obtained with PSF and "aperture" photometry. These differences are due to the
other sources influencing the background determination. In addition, there is a
lot of diffuse background emission present, which can be seen in
Fig.\,\ref{hh211_data} as the difference between the PHOT photometry and the
LWS continuum. So, it is very difficult to determine the background and to
state which pixel contributes to the flux of which object. This is further
complicated by the fact that HH\,211\,FIRS2 has a higher surface brightness
than our point source HH\,211-MM at 160 and 200\,$\mu$m. Additionally,
IC\,348\,IR could influence our measured flux for HH\,211-MM also.

Due to the difficulties in the determination of the fluxes of HH\,211-MM we
supplemented the ISOPHOT data with SCUBA datapoints at 450 and 850\,$\mu$m from
Rengel et al. \cite{rfeh01} and JCMT bolometry at 1.1\,mm from McCaughrean et
al. \cite{mrz94}. The fluxes are given in Table\,\ref{phot_fluxes}. We find a
fit ($rms$\,=\,2.0) with the following object parameters: T\,=\,21.0\,K,
$\beta$\,=\,1.5, and $\Sigma \Omega$\,=\,12.5$\times$10$^{-10}$\,sr. In this
case a variable optical depth (instead of a fixed value of unity) does not
improve the fit. The inferred bolometric temperature is 31.4\,K and the
bolometric luminosity is 4.5\,L$_\odot$, while 4.5\% account for L$_{smm}$.
This verifies the Class\,0 nature of HH\,211-MM. According to the low
bolometric temperature we find a high $\tau_{100}$ on the order of 20 and a
large radius for the $\tau_{100} = 1$ surface of 520\,AU.

\subsection{IC\,1396\,W}

\begin{figure}
\includegraphics[bb=140 365 370 510]{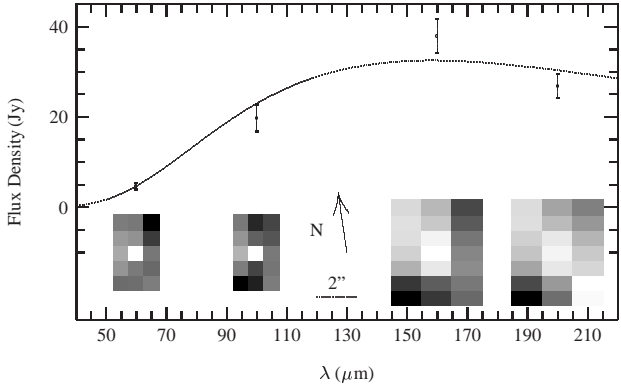}
\caption{\label{ic1396w_data} As Fig.\,\ref{cepe_data}, but for
IC\,1396\,W. We do not have LWS data for this source. The best graybody fit has
a temperature of T\,=\,30.0\,K, $\beta$\,=\,0.3, and $\Sigma
\Omega$\,=\,1.1$\times$10$^{-10}$\,sr. $\tau_{100}$ was fixed to unity.
Extended cool dust, or a close group of cold sources, are seen northeast of
IC\,1396\,W, while a very cold bright source appears at 200\,$\mu$m to the
south-west.}
\end{figure}

Our PHOT maps and the derived photometry for IC\,1396\,W are shown in
Fig.\,\ref{ic1396w_data}, together with the graybody fit to these data. In the
maps, two additional sources are evident. IC\,1396\,W\,FIRS2, to the north-east
of IC\,1396\,W, peaks at 160\,$\mu$m, whereas IC\,1396\,W\,FIRS3, to the
south-west, is remarkably red, with a flux ratio at C$_{200}$/C$_{160}$ of 2.6.
These additional objects were not detected with the C100 detector, because the
maps at these wavelengths are slightly smaller and the objects just fall
outside. Nevertheless, they influence the flux measurements, especially in the
C200 wavelengths range. We do not have observations at longer (sub-millimeter
or millimeter) wavelengths to supplement the ISO data. 

For IC\,1396\,W we find a temperature of T\,=\,30.0\,K, $\beta$\,=\,0.3, and
$\Sigma \Omega$\,=\,1.1$\times$10$^{-10}$\,sr for the best graybody fit. These
parameters are obtained fixing $\tau_{100}$ to unity. A variable optical depth
does not improve the fit. The fit is poor ($rms$\,=\,1.3), however, suggesting
errors in the photometry or a source with dust at more than one temperature.
With the graybody parameters we determine a bolometric temperature of 32.6\,K,
a luminosity of 16.4\,L$_\odot$, and a L$_{smm}$/L$_{bol}$ ratio of 0.059.
Concerning the difficulties in the flux measurements and the not available
observations at longer wavelengths, the classification of IC\,1396\,W as a
Class\,0 source remains questionable. At least sub-mm observations are needed
to confirm the presence of a spatially extended envelope, to ensure that we do
not just see a compact disk edge-on.

\subsection{L\,1211}

\label{res_l1211}

\begin{figure}
\includegraphics[bb=140 365 370 510]{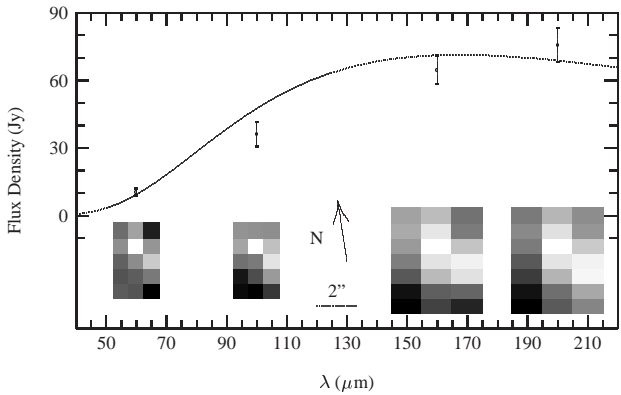}
\caption{\label{l1211_data} As Fig.\,\ref{ic1396w_data}, but for L\,1211.
The best fit is a blackbody with T\,=\,30.5\,K, and $\Sigma
\Omega$\,=\,2.1$\times$10$^{-10}$\,sr. $\tau_{100}$ was fixed to unity. The
object appears north of the nominal IRAS position, and a second cool source is
detected south-west of it. See text for the identification of these sources.}
\end{figure}
 
Our PHOT maps of L\,1211, the derived photometry, and a graybody fit to these
data are displayed in Fig.\,\ref{l1211_data}. Somewhat to our surprise, the
L\,1211 source was not found at its nominal IRAS position, but is shifted a
full pixel, corresponding to 45\arcsec\, to the north. A second source
L\,1211\,FIRS2, is found in the south-west. Comparing our maps with the work of
Tafalla et al. \cite{tmmb99} and Anglada \& Rodr\'{\i}guez \cite{ar02} we find
that our object L\,1211 is identical to MMS\,4 or VLA\,5 and the object
L\,1211\,FIRS2 seems to be a superposition of MMS\,3, MMS\,2, and MMS\,1, and
VLA\,3 and VLA\,1, respectively. 

The best fits to both sources are blackbodies ($\beta$\,=\,0.0) and have
temperatures of T\,=\,30.5\,K and T\,=\,26.8\,K for L\,1211 and L\,1211\,FIRS2,
respectively. We infer a solid angle $\Sigma \Omega$ of the sources of 2.1 and
1.7$\times$10$^{-10}$\,sr. This results in bolometric temperatures of 30.5 and
26.9\,K and in bolometric luminosities of 33.1 and 16.0\,L$_\odot$, for L\,1211
and L\,1211\,FIRS2, respectively. Even if we get quite high L$_{smm}$/L$_{bol}$
ratios (0.073 and 0.100), we cannot firmly establish the classification of the
sources as Class\,0 objects. This is because our ISOPHOT data do not cover the
emission maximum and also L\,1211\,FIRS2 is a superposition of the emission
from several sources.

Tafalla et al. \cite{tmmb99} classify L\,1211 as a transitional object between
Class\,0 and Class\,1. They use the IRAS fluxes and an additional observation
at 1.2\,mm (see Table\,\ref{phot_fluxes}). With these data, as for our ISO
data, the maximum of the emission could not be determined exactly. It could
only be constrained to lie between 100 and 1200\,$\mu$m. With our ISO data, we
could corroborate the assumption that L\,1211 is of Class\,0, since the maximum
of the SED is at $\lambda > 160$\,$\mu$m. If we use the 1.2\,mm datapoint from
Tafalla et al. \cite{tmmb99} for our analysis, we get a poor fit. Using both,
the ISOPHOT and the 1.2\,mm point, with plausible values for $\beta$
(1.0..2.0), the 1.2\,mm flux is always overestimated by about one order of
magnitude. The same applies for L\,1211\,FIRS2. Accepting values of
$\beta$\,=\,3.0 or higher, we determine T\,=\,36.3, $\Sigma
\Omega$\,=\,3.0$\times$10$^{-10}$\,sr (L\,1211) and T\,=\,24.8, $\Sigma
\Omega$\,=\,5.8$\times$10$^{-10}$\,sr (L\,1211\,FIRS2). This leads to
bolometric temperatures of about 90 and 46\,K and bolometric luminosities of 77
and 19\,L$_\odot$ for L\,1211 and L\,1211\,FIRS2, respectively. The resulting
L$_{smm}$/L$_{bol}$ ratios of 0.001 and 0.004 would then classify both objects
as Class\,1 (L\,1211\,FIRS2 is still near the transition phase between Class\,0
and Class\,1).

\subsection{RNO\,15\,FIR}

\label{res_rno15fir}

\begin{figure}
\includegraphics[bb=140 365 370 510]{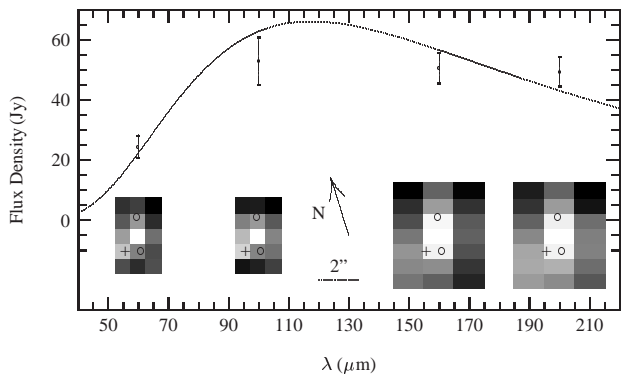}
\caption{\label{rno15fir_data}
As Fig.\,\ref{ic1396w_data}, but for RNO\,15\,FIR. The best fit has temperature
of T\,=\,34.0\,K, $\beta$\,=\,1.1, and $\Sigma
\Omega$\,=\,0.7$\times$10$^{-10}$\,sr. $\tau_{100}$ was fixed to unity.
South-east of RNO\,15\,FIR, the warmer source RNO\,15 is detected as well
(marked by a cross), especially at the shorter wavelengths. The positions of
two other weak sub-mm sources (SMS1 -- north, SMS2 -- south) are indicated by a
circle (Rengel at al. (2001, 2002)).
}
\end{figure}
 
The ISOPHOT maps of the RNO\,15\,FIR region are shown in
Fig.\,\ref{rno15fir_data}, together with the derived photometry and the
graybody fit. Visible on our maps are RNO\,15\,FIR in the centre, and the
source RNO\,15 to the south-east, marked by a cross. Since this source is
warmer than RNO\,15\,FIR, it is prominent at the shorter wavelengths, but fades
considerably relative to RNO\,15\,FIR towards the longer wavelengths (due to
the larger pixel size the two objects also merge). From higher spatial
resolution sub-mm maps at 450 and 850\,$\mu$m taken with SCUBA (Rengel et al.
\cite{rfhe02}) we know that two other objects SMS1 and SMS2 are present to the
north and south of RNO\,15\,FIR, but are merged with it at ISOPHOT resolution
(marked by a circle in Fig.\,\ref{rno15fir_data}). They surely influence our
flux measurements of RNO\,15\,FIR.

Since the ISOPHOT measurements show a broad and not well defined maximum of the
SED, we supplement these data with SCUBA measurements of Rengel et al.
\cite{rfeh01} to determine more accurate source properties. The fluxes at 450
and 850\,$\mu$m are measured in a 45\arcsec\, by 45\arcsec\, aperture (Rengel
priv. communication) and listed in Table\,\ref{phot_fluxes}. Using these data
in combination with the ISOPHOT points, we get T\,=\,34.0\,K, $\beta$\,=\,1.1,
and $\Sigma \Omega$\,=\,1.7$\times$10$^{-10}$\,sr with an optical depth
$\tau_{100}$ fixed to unity. The resulting bolometric temperature and
luminosity are 44.6\,K and 8.4\,L$_\odot$. 1.7\% of the luminosity is in the
sub-millimeter regime, classifying RNO\,15\,FIR as a Class\,0 source. If we
vary the optical depth at 100\,$\mu$m also, the fit is improved (the $rms$ is
lowered from 1.0 to 0.6), but the parameters T$_{bol}$ and L$_{bol}$ do not
change. 

Davis et al. \cite{drec97} suggested that RNO\,15\,FIR might be a double
source, due to the observed wiggling in the outflow. This might also be
indicated by the deviation of the data points from the determined SED (see
Fig.\,\ref{scuba_data}).

%%%%%%%%%%%%%%%%%%%%%%%%%%%%%%%%%%%%%%%%%%%%%%%%%%%%%%%%%%%%%%%%%%%%%%%%%%%%%%
%%%%%%%%%%%%%%%%%%%%%%   Models        %%%%%%%%%%%%%%%%%%%%%%%%%%%%%%%%%%%%%%%
%%%%%%%%%%%%%%%%%%%%%%%%%%%%%%%%%%%%%%%%%%%%%%%%%%%%%%%%%%%%%%%%%%%%%%%%%%%%%%

\section{Discussion}

\label{discussion}

\subsection{Mass and Age determination}

\begin{figure}
\includegraphics[width=8.5cm, bb=10 10 515 427]{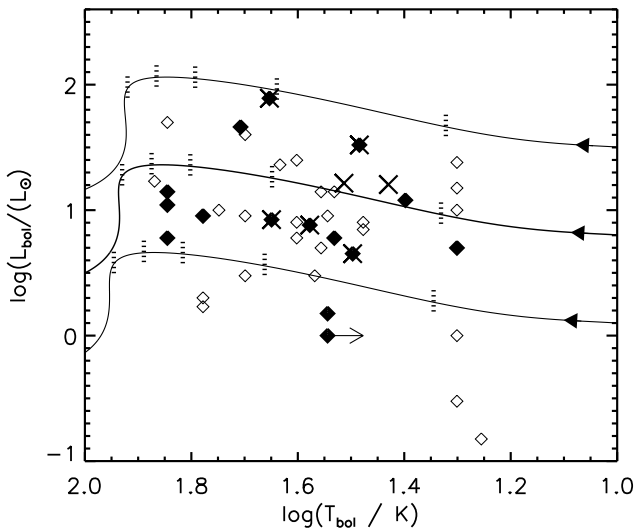}
\caption{\label{classzero}
The bolometric luminosity--temperature diagram for the objects analysed here
(thick Xs), the Class\,0 data from Table\,\ref{lbolvslhtwo} (filled diamonds),
and the Class\,0 data from the review of Andr\'e et al. (2000) (open
diamonds). The superimposed evolutionary  tracks are discussed in
Section\,\ref{evolscheme}. Protostars evolve from right to left. Three tracks
for final masses of 0.2, 1, and 5\,M$_\odot$ are displayed. The model peak 
accretion rate is reached at 17,000\,yr, and the power law fall-off is
$\propto$\,t$^{-7/4}$ with time t, on a 30,000\,yr timescale. The vertical
dotted lines on the tracks mark the model ages of 20, 30, 40, 50, and 75
thousand years.
}
\end{figure}

Do Class\,0 objects develop into Class\,1 and Class\,2 protostars? To answer
this, we wish to determine basic parameters for the Class\,0 protostars, such
as age, surrounding mass, present mass and final mass. These, however, are
model dependent quantities. In Fig.\,\ref{classzero} we plot the locations of
our seven (including L\,1211\,FIRS2) sources on the L$_{bol}$\,--\,T$_{bol}$
diagram (large crosses), which is the protostellar equivalent of a
Hertzsprung-Russell diagram (Myers et al. \cite{macs98}). Also plotted on the
diagram are the data for another 37 Class\,0 protostars, as listed by Andr{\'e}
et al. \cite{awb00}. Two of these sources possess bolometric luminosities above
1000\,L$_\odot$, and  so fall outside the display. Note that Class\,0
protostars possess bolometric temperatures below  $\sim 80$\,K.

The present sample contains quite powerful and cold Class\,0 members. Two
sources lie above the location of the other explored sources. These are
L\,1211 and Cep\,E. As we demonstrate below, such powerful Class\,0 sources
with low bolometric temperature, can indeed be included in an evolutionary
model through the Classes\,0\,--\,1\,--\,2. The large surrounding
masses observed restrict the type of model and these objects
could go on to produce high-mass stars.

The model tracks plotted represent the evolution of three protostars which end
up accumulating masses of 0.2, 1, and 5\,M$_\odot$. The tracks were derived by
combining the Unification Scheme, as reviewed by Smith \cite{s00,s02}, with the
framework for protostellar envelopes presented by Myers et al. \cite{macs98},
according to the prescription presented below. We thus determine model ages,
present masses of the protostellar nucleus, envelope masses and the final
stellar masses (Table\,\ref{parameters}). The result is that the more massive
Class\,0 protostars possess large envelopes and would become massive stars.
According to the model described here, most of the envelope, however, is not
accreted but dispersed, if the majority of protostars here are to form low-mass
stars. Note that alternative schemes have been presented by Bontemps et al.
\cite{batc96}, Saraceno et al. \cite{saraceno96}, and Andr\'e et al.
\cite{awb00}. The model envelope masses are in agreement with the measurements.
Just for the two objects where we could not determine a proper bolometric
temperature (L\,1211 and L\,1211\,FIRS2), there is a significant deviation.

\subsection{Outflow Luminosity vs. Source Properties}

\begin{table}
\renewcommand{\tabcolsep}{2.5pt}
\caption{\label{lbolvslhtwo}
Summary of the Class\,0 sources for which a correlation of the source
properties with the outflow luminosity in the 1\,--\,0\,S(1) line of H$_2$ was
investigated. Except for our objects observed with ISOPHOT, L$_{bol}$ and
T$_{bol}$ are adapted from Andr\'e et al. (2000), as well as all the
M$_{env}$ values (except RNO\,15\,FIR and L\,1211). The outflow luminosities
are either from published literature or our own measurements. The typical
errors of our outflow luminosities are 10\%. In the Ref. column the references
are given where we took L$_{\mbox{\tiny H$_2$\,1\,--\,0\,S(1)}}$ measurements
from.
}
\begin{center}
\begin{tabular}{lccccl}
Object & L$_{bol}$ & T$_{bol}$ & M$_{env}$ & L$_{\mbox{\tiny
H$_2$\,1\,--\,0\,S(1)}}$ & Ref.\\  
& [L$_\odot$] & [K] & [M$_\odot$] & [10$^{-3}$\,L$_\odot$] & \\
\noalign{\smallskip}
\hline
\noalign{\smallskip}
RNO\,15\,FIR                & ~~8.4 & 44.6        & ~~0.9$^*$      & 0.46     & 1, 7 \\
HH\,211-MM                  & ~~4.5 & 31.4        & 1.5            & 3.1~~    & 2, 7 \\      
L\,1157                     & ~~7.6 & 37.8        & 0.5            & 6.1~~    & 3, 7 \\
IC\,1396\,W                 & 16.4  & 32.6        & --             & 19.0~~~~ & 6 \\
L\,1211                     & 33.1  & 30.5        & ~~~~0.8$^{**}$ & 10.7~~~~ & 7 \\
Cep\,E                      & 77.9  & 45.0        & 7.0            & 70.0~~~~ & 4, 7 \\ 
\noalign{\smallskip} \hline \noalign{\smallskip}
L\,1448\,N         & 11.0  & 70.0        & 2.3            & 2.16     & 7 \\
L\,1448\,IRS2      & ~~6.0 & 70\,?~      & 0.9            & 2.8~~    & 7 \\
L\,1448\,C         & ~~9.0 & 60.0        & 1.4            & 5.7~~    & 7 \\
IRAS\,03282        & ~~1.5 & 35.0        & 0.6            & 4.46     & 7 \\
HH\,212\,MM        & 14.0  & 70\,?~      & 1.2            & 5.3      & 5, 7 \\
HH\,24\,MMS        & ~~5.0 & 20\,?~      & 4.0            & 1.21     & 1, 7 \\
HH\,25\,MMS        & ~~6.0 & 34.0        & 0.5            & 6.61     & 1 \\
NGC\,2264\,G\,VLA2 & 12.0  & 25.0        & 2.0            & 7.75     & 3 \\
VLA\,1623          & ~~1.0 & $<$35~~~~~~ & 0.7            & 0.81     & 3, 7 \\
Ser\,--\,FIRS1     & 46.0  & 51.0        & 3.0            & 0.64     & 7 \\
\noalign{\smallskip}
\hline
\noalign{\smallskip}
\end{tabular}
\end{center}
\begin{list}{}{}
 \item[*] taken from Rengel et al. \cite{reh03}
 \item[**] taken from Tafalla et al. \cite{tmmb99}
\end{list}
{\bf References:} (1) Davis et al. \cite{drec97} (2) McCaughrean et al.
\cite{mrz94} (3) Davis \& Eisl\"offel \cite{de95} (4) Eisl\"offel et al.
\cite{esdr96} (5) Zinnecker et al. \cite{zmr98} (6) Froebrich \& Scholz
\cite{fs03} (7) own measurements
\end{table}

Is the luminosity of the outflows from the Class\,0 sources correlated with the
properties of the sources like their bolometric luminosity, the temperature, or
mass of their envelopes? To answer this question we measured the luminosity of
the outflows from 16 of the Class\,0 sources in Andr{\'e} et al. \cite{awb00}
and our objects, in the 1\,--\,0\,S(1) line of molecular hydrogen at
2.122\,$\mu$m. Due to the short cooling times (some years, Smith and Brand
\cite{sb90}), H$_2$ is a good tracer of emission of shocked gas caused by
current interactions between outflowing material and the surrounding gas. The
1\,--\,0\,S(1) line of H$_2$ is usually the brightest ro-vibrational line in a
spectrum of shocked gas and thus most easily detected. 

\begin{figure}
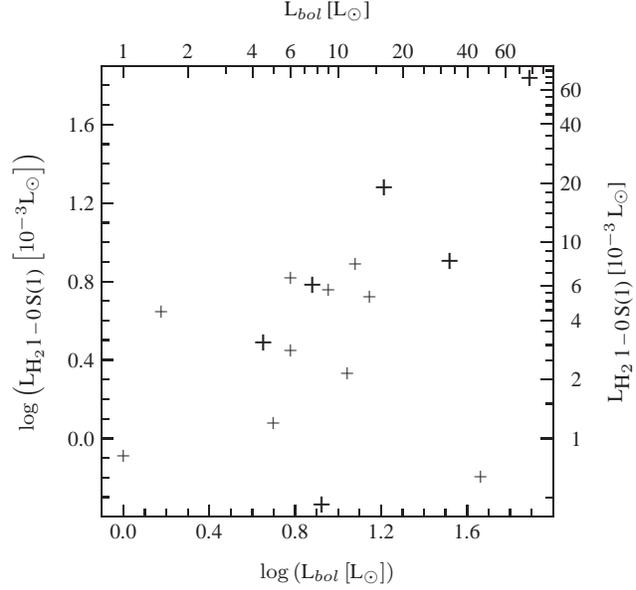

\beginpicture
\setcoordinatesystem units <28.571mm,26.087mm> point at 0 0
\setplotarea x from -0.10 to 2.0, y from -0.4 to 1.90
\small
\axis left label
{\begin{sideways} $\log \left( \mbox{L}_{\mbox{H}_2 \mbox{1\,--\,0\,}
\mbox{S(1)}} \left[ 10^{-3} \mbox{L}_\odot \right] \right) $ \end{sideways}}
ticks in long numbered from -0. to 1.8 by 0.4 
    in short unlabeled from -0.4 to 1.8 by 0.1 /
\axis right label
{\begin{sideways}L$_{\mbox{H$_2$\,1\,--\,0\,S(1)}}$\,[10$^{-3}$\,L$_\odot$]\end{sideways}}
ticks in long logged numbered from 1 to 1 by 1 
                           from 2 to 6 by 2 
                           from 10 to 20 by 10 
                           from 40 to 60 by 20 
    in short logged unlabeled from 0.5 to 6 by 0.5 
                           from 6 to 10 by 1 
                           from 10 to 20 by 2 
                           from 20 to 75 by 5 /
\axis bottom label {$\log \left( \mbox{L}_{bol} \left[ \mbox{L}_\odot \right]
\right)$}
ticks in long numbered from 0 to 1.9 by 0.4
    in short unlabeled from 0 to 1.9 by 0.1 /
\axis top label {L$_{bol}$\,[L$_\odot$]}
ticks in long logged numbered from 1 to 1 by 1 
                              from 2 to 6 by 2
                              from 10 to 10 by 10
                              from 20 to 60 by 20
    in short logged unlabeled from 1 to 10 by 1
                              from 10 to 40 by 5 
                              from 40 to 90 by 10 /
                           
\put {{\Large \bf +}} at 0.9243 -0.33724
\put {{\Large \bf +}} at 0.6532 0.49136
\put {{\Large \bf +}} at 0.8808 0.78533
\put {{\Large \bf +}} at 1.2148 1.27875
\put {{\Large \bf +}} at 1.5198 0.90309
\put {{\Large \bf +}} at 1.8915 1.83885
%l1448N
\put {{\normalsize +}} at 1.04139 0.33445
%l1448IRS2
\put {{\normalsize +}} at 0.77815 0.44715
%l1448C
\put {{\normalsize +}} at 0.95424 0.75587
%IRAS03282
\put {{\normalsize +}} at 0.17609 0.6492988
%HH212MM
\put {{\normalsize +}} at 1.14613 0.724276
%HH24MMS
\put {{\normalsize +}} at 0.69897 0.081099
%HH25MMS
\put {{\normalsize +}} at 0.77815 0.82023
%VLA1623
\put {{\normalsize +}} at 0.00000 -0.08951
%NGC2264G VLA2
\put {{\normalsize +}} at 1.07918 0.88899
%Ser FIRS1
\put {{\normalsize +}} at 1.66276 -0.19525
                                                 
\endpicture
\caption{\label{outflows_lbol} Measured outflow luminosity in the
1\,--\,0\,S(1) line of H$_2$ versus the bolometric source luminosity for the
sources listed in Table\,\ref{lbolvslhtwo}. The objects investigated in this
paper are marked with a large $+$ sign. No significant correlation is found for
this sample.}
\end{figure}

\begin{figure}
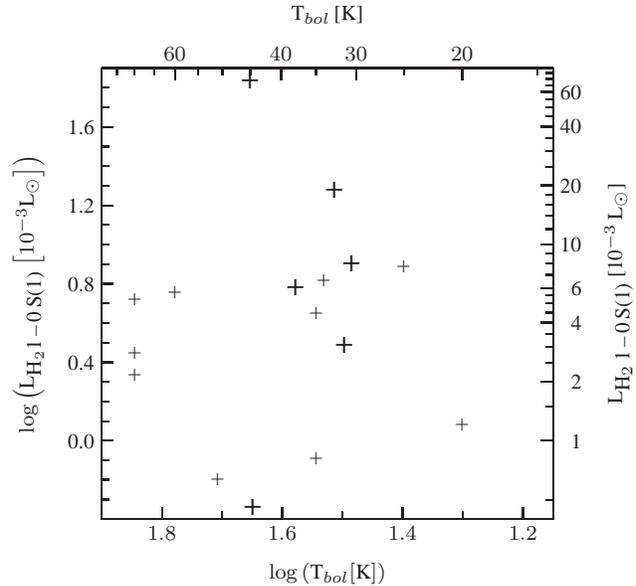

\beginpicture
\setcoordinatesystem units <-80mm,26.087mm> point at -1 0
\setplotarea x from 1.9 to 1.15, y from -0.4 to 1.90
\small
\axis left label
{\begin{sideways} $\log \left( \mbox{L}_{\mbox{H}_2 \mbox{1\,--\,0\,}
\mbox{S(1)}} \left[ 10^{-3} \mbox{L}_\odot \right] \right) $ \end{sideways}}
ticks in long numbered from -0. to 1.8 by 0.4 
    in short unlabeled from -0.4 to 1.8 by 0.1 /
\axis right label
{\begin{sideways}L$_{\mbox{H$_2$\,1\,--\,0\,S(1)}}$\,[10$^{-3}$\,L$_\odot$]\end{sideways}}
ticks in long logged numbered from 1 to 1 by 1 
                           from 2 to 6 by 2 
                           from 10 to 20 by 10 
                           from 40 to 60 by 20 
    in short logged unlabeled from 0.5 to 6 by 0.5 
                           from 6 to 10 by 1 
                           from 10 to 20 by 2 
                           from 20 to 75 by 5 /
\axis bottom label {$\log \left( \mbox{T}_{bol} [\mbox{K}] \right)$}
ticks in long numbered from 1.2 to 1.9 by 0.2
    in short unlabeled from 1.2 to 1.9 by 0.1 /
\axis top label {T$_{bol}$\,[K]}
ticks in long logged numbered from 20 to 40 by 10 
                              from 60 to 60 by 20
    in short logged unlabeled from 15 to 75 by 5  /
                           
\put {{\Large \bf +}} at 1.6493 -0.33724
\put {{\Large \bf +}} at 1.4969 0.49136
\put {{\Large \bf +}} at 1.5775 0.78533
\put {{\Large \bf +}} at 1.5132 1.27875
\put {{\Large \bf +}} at 1.4843 0.90309
\put {{\Large \bf +}} at 1.6532 1.83885
%l1448N
\put {{\normalsize +}} at 1.84510 0.33445
%l1448IRS2
\put {{\normalsize +}} at 1.84510 0.44715
%l1448C
\put {{\normalsize +}} at 1.77815 0.75587
%IRAS03282
\put {{\normalsize +}} at 1.54407 0.6492988
%HH212MM
\put {{\normalsize +}} at 1.84510 0.724276
%HH24MMS
\put {{\normalsize +}} at 1.30103 0.081099
%HH25MMS
\put {{\normalsize +}} at 1.53148 0.82023
%VLA1623
\put {{\normalsize +}} at 1.54407 -0.08951
%NGC2264G VLA2
\put {{\normalsize +}} at 1.39794 0.88899
%Ser FIRS1
\put {{\normalsize +}} at 1.70757 -0.19525
                                                 
\endpicture
\caption{\label{outflows_tbol} As Fig.\,\ref{outflows_lbol} but for the
bolometric temperatures of the Class\,0 sources from Table\,\ref{lbolvslhtwo}. 
No significant correlation is found for this sample.} 
\end{figure}

\begin{figure}
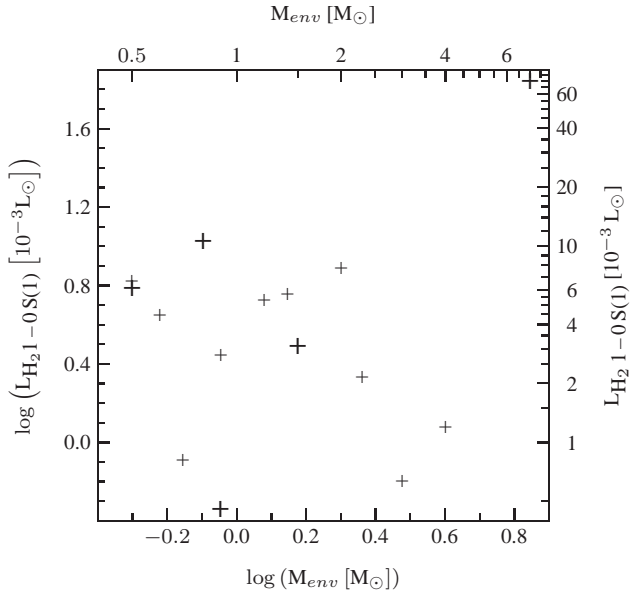

\beginpicture
\setcoordinatesystem units <46.154mm,26.087mm> point at 0 0
\setplotarea x from -0.40 to 0.9, y from -0.4 to 1.9
\small
\axis left label
{\begin{sideways} $\log \left( \mbox{L}_{\mbox{H}_2 \mbox{1\,--\,0\,}
\mbox{S(1)}} \left[ 10^{-3} \mbox{L}_\odot \right] \right) $ \end{sideways}}
ticks in long numbered from -0. to 1.8 by 0.4 
    in short unlabeled from -0.4 to 1.8 by 0.1 /
\axis right label
{\begin{sideways}L$_{\mbox{H$_2$\,1\,--\,0\,S(1)}}$\,[10$^{-3}$\,L$_\odot$]\end{sideways}}
ticks in long logged numbered from 1 to 1 by 1 
                           from 2 to 6 by 2 
                           from 10 to 20 by 10 
                           from 40 to 60 by 20 
    in short logged unlabeled from 0.5 to 6 by 0.5 
                           from 6 to 10 by 1 
                           from 10 to 20 by 2 
                           from 20 to 75 by 5 /

\axis bottom label {$\log \left( \mbox{M}_{env} \left[ \mbox{M}_\odot \right]
\right)$}
ticks in long numbered from -0.2 to 0.8 by 0.2
    in short unlabeled from -0.4 to 0.9 by 0.1 /
\axis top label {M$_{env}$\,[M$_\odot$]}
ticks in long logged numbered from 2 to 6 by 2 
                              from 0.5 to 0.5 by 0.5
                              from 1 to 1 by 1
    in short logged unlabeled from 1 to 7.5 by 0.5  /
                          
%RNO15FIR
\put {{\Large \bf +}} at -0.04576 -0.33724
%HH211
\put {{\Large \bf +}} at 0.17609 0.49136
%L1157
\put {{\Large \bf +}} at -0.30103 0.78533
%L1211
\put {{\Large \bf +}} at -0.09691 1.02938
%CepE
\put {{\Large \bf +}} at 0.84510 1.84501

%l1448N
\put {{\normalsize +}} at 0.36173 0.33445
%l1448IRS2
\put {{\normalsize +}} at -0.04576 0.44715
%l1448C
\put {{\normalsize +}} at 0.14613 0.75587
%IRAS03282
\put {{\normalsize +}} at -0.22185 0.6492988
%HH212MM
\put {{\normalsize +}} at 0.07918 0.724276
%HH24MMS
\put {{\normalsize +}} at 0.60206 0.081099
%HH25MMS
\put {{\normalsize +}} at -0.30103 0.82023
%VLA1623
\put {{\normalsize +}} at -0.15490 -0.08951
%NGC2264G VLA2
\put {{\normalsize +}} at 0.30103 0.88899
%Ser FIRS1
\put {{\normalsize +}} at 0.47712 -0.19525
                                                 
\endpicture
\caption{\label{outflows_menv} As Fig.\,\ref{outflows_lbol} but for the
envelope masses of the Class\,0 sources from Table\,\ref{lbolvslhtwo}. For
IC\,1396\,W we do not have measurements of the envelope mass. No
significant correlation is found for this sample.} 
\end{figure}

In magnetohydrodynamic models of Class\,0 sources (e.g. Shu et al.
\cite{snowrl94}, Hirose et al. \cite{husm97}, and Ouyed and Pudritz
\cite{op97}) the accretion rate onto the protostar is connected to the amount
of material injected into the outflowing jet. This material interacts with the
surrounding quiescent gas in shocks. Hence, the luminosity of these shocks may
be connected to the mass accretion rate and thus to the source properties.
Here, we tested for correlations of the 1\,--\,0\,S(1) H$_2$ luminosity with
the source bolometric luminosity, the bolometric temperature and the mass of
the surrounding protostellar envelope given by Andr{\'e} et al. \cite{awb00}.
The results of these comparisons are shown in Figs.\,\ref{outflows_lbol},
\ref{outflows_tbol}, and \ref{outflows_menv}. We obtained a linear regression
for each case and tested if the slope of the regression line differed
statistically significantly from a slope value of zero. With a probability of
error of 5\,\% none of the regression lines differs from a constant value.
Additionally a Kolmogorow-Smirnow-Test shows, that with a probability of error
of 0.1\,\% the data is not consistent with a constant value. Thus, a
significant correlation of the outflow luminosity in the 1\,--\,0\,S(1) line of
H$_2$ with any source parameter was not found.

%%%%%%%%%%%%%%%%%%%%%%%%%%%%%%%%%%%%%%%%%%%%%%%%%%%%%%%%%%%%%%%%%%%%%%%%%%%%%%
%%%%%%%%%%%%%%%%%%%%%%   Outflows    %%%%%%%%%%%%%%%%%%%%%%%%%%%%%%%%%%%%%%%%%
%%%%%%%%%%%%%%%%%%%%%%   lbol        %%%%%%%%%%%%%%%%%%%%%%%%%%%%%%%%%%%%%%%%%
%%%%%%%%%%%%%%%%%%%%%%%%%%%%%%%%%%%%%%%%%%%%%%%%%%%%%%%%%%%%%%%%%%%%%%%%%%%%%%

%%%%%%%%%%%%%%%%%%%%%%%%%%%%%%%%%%%%%%%%%%%%%%%%%%%%%%%%%%%%%%%%%%%%%%%%%%%%%%
%%%%%%%%%%%%%%%%%%%%%%   Outflows    %%%%%%%%%%%%%%%%%%%%%%%%%%%%%%%%%%%%%%%%%
%%%%%%%%%%%%%%%%%%%%%%   tbol        %%%%%%%%%%%%%%%%%%%%%%%%%%%%%%%%%%%%%%%%%
%%%%%%%%%%%%%%%%%%%%%%%%%%%%%%%%%%%%%%%%%%%%%%%%%%%%%%%%%%%%%%%%%%%%%%%%%%%%%%

%%%%%%%%%%%%%%%%%%%%%%%%%%%%%%%%%%%%%%%%%%%%%%%%%%%%%%%%%%%%%%%%%%%%%%%%%%%%%%
%%%%%%%%%%%%%%%%%%%%%%   Outflows    %%%%%%%%%%%%%%%%%%%%%%%%%%%%%%%%%%%%%%%%%
%%%%%%%%%%%%%%%%%%%%%%   menv        %%%%%%%%%%%%%%%%%%%%%%%%%%%%%%%%%%%%%%%%%
%%%%%%%%%%%%%%%%%%%%%%%%%%%%%%%%%%%%%%%%%%%%%%%%%%%%%%%%%%%%%%%%%%%%%%%%%%%%%%

The lack of such correlations may have various explanations. For example, in
each outflow, we observe H$_2$ emission at various distances from the source
and these knots or bow shocks are indicating material which was ejected from
the source at different times in the past ($\Delta t$ = distance to the
source/jet velocity). Also, the knot luminosity depends on the local properties
of the surrounding gas (e.g. gas density, atomic fraction). Additionally, the
extinction gradient in the K-band along the outflow is not known. It will alter
the measured relative and total fluxes in the sense that knots closer to the
source appear fainter due to higher extinction.  In Section~\ref{evolscheme},
however, we argue that the location on these diagrams depends sensitively on
both mass and age, which results in a wide scatter. Therefore, the lack of  a
significant correlation of the present source properties and the outflow
luminosity in the H$_2$ 1\,--\,0\,S(1) line may not be surprising. 

A better tool for comparing the outflows to source properties may be an
optically thin line of CO (e.g. the 1\,--\,0\,$^{13}$CO line), which should
give a measurement of the time-integrated power of the outflow without being
influenced by local extinction effects. Comparable observations in the same
transition and isotope of CO are needed for a statistically reasonable sample
of objects to study their behaviour. At present, only small samples of Class\,0
sources have been thus analysed (e.g. Bontemps et al. \cite{batc96}, Smith
\cite{s00,s02}).

%%%%%%%%%%%%%%%%%%%%%%%%%%%%%%%%%%%%%%%%%%%%%%%%%%%%%%%%%%%%%%%%%%%%%%%%%%%%%%
%%%%%%%%%%%%%%%%%%%%%%   Modelling    %%%%%%%%%%%%%%%%%%%%%%%%%%%%%%%%%%%%%%
%%%%%%%%%%%%%%%%%%%%%%%%%%%%%%%%%%%%%%%%%%%%%%%%%%%%%%%%%%%%%%%%%%%%%%%%%%%%%%

\section{An evolutionary scheme}

\label{evolscheme}

An evolutionary model for protostars is presented in the Appendix.  The outflow
scheme has been elaborated by Smith \cite{s98,s00,s02} and applied by Davis et
al. \cite{dsm98}, Yu et al. \cite{ybsbb00}, and Stanke et al. \cite{stanke00}.
It is based on a prescribed accretion rate from an envelope. Modelling of 
outflows has demonstrated that the fraction of mass which escapes through jets
must reach a maximum during the Class\,0 stage. This is required to account for
the excess momentum and power of Class\,0 bipolar outflows, as calculated from
observations of emission lines of CO rotational transitions (see Smith
\cite{s00}). We outline in the Appendix the  fundamental formulae of the
evolutionary scheme.

According to previous modelling of the envelope, three parameters must be 
introduced to generate plausible models for the bolometric temperature. As
shown by  Myers et al. \cite{macs98}, these are (1) the envelope's outer
temperature (here T$_o$\,=\,24\,K), (2) the efficiency of accretion of the
envelope into the star-jet system  and (3) the difference in  evolutionary
timescale between the envelope and the protostar. The envelope consists of
material which will fall onto the central object as well as mass directly lost
soon  after the Class\,0 stage. This extra mass component proves necessary to
produce a low bolometric temperature, as observed for the Class\,0 sources, yet
must be rapidly lost in order to yield T\,Tauri stars within a reasonable time
(Myers et al. \cite{macs98}).  

\begin{figure}
\includegraphics[width=8.5cm, bb=15 10 515 428]{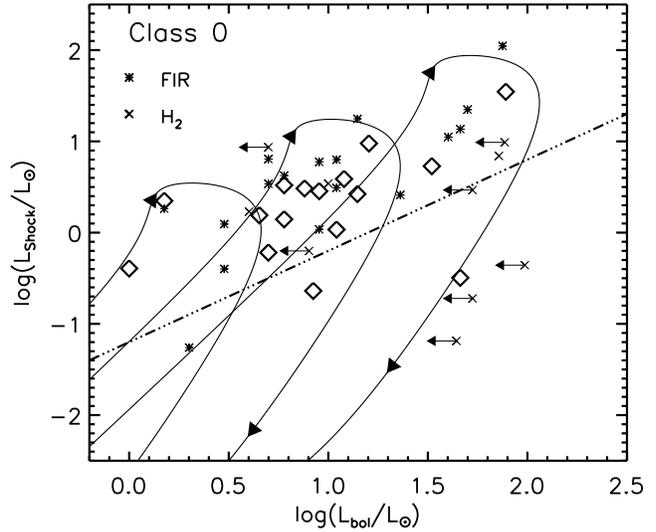}
\caption{\label{h2-lbol0}
The derived outflow shock luminosity versus the bolometric  source luminosity
for the Class\,0 sources listed in Table\,\ref{lbolvslhtwo}, as well  as the
Class\,0 sample investigated by Stanke (2000) (symbol: `x') and the
far-infrared line  ISO luminosities presented by Giannini et al. (2001)
(symbol: '*'). The model  tracks are for the same three models presented in
Fig\,\ref{classzero} and the  straight line divides model Class\,0 and model
Class\,1 protostars, as determined by the protostar possessing half of its
final mass.
}
\end{figure}

\begin{figure}
\includegraphics[width=8.5cm, bb=15 10 515 428]{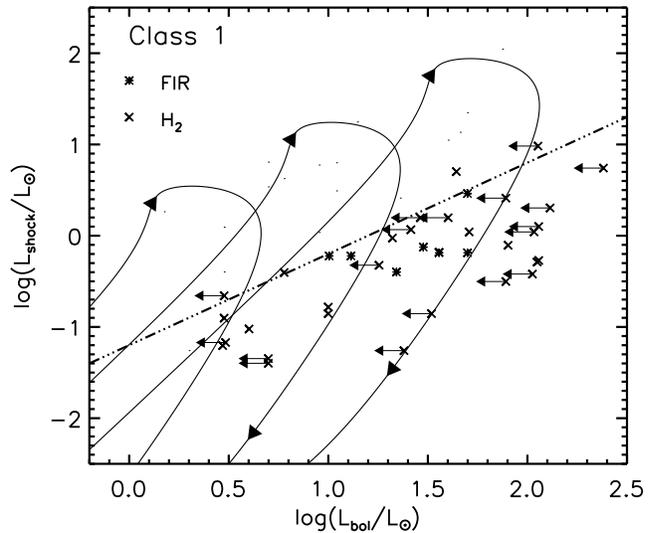}
\caption{\label{h2-lbol1}
The derived outflow shock luminosity versus the bolometric source luminosity
for Class\,1 sources from the sample of Stanke (2000) (symbol: `x') 
and the far-infrared line ISO luminosities presented by Giannini et al.
(2001) (symbol: '*'). The model tracks are for the same three models
presented in Fig\,\ref{classzero} and the straight line divides model Class\,0
and model Class\,1 protostars, as determined by the protostar possessing half
of its final mass.
}
\end{figure}

Previously, we modelled the envelope evolution by assuming mass conservation. 
Here, we find two significant adjustments are necessary in order to model the 
new data and maintain plausible time scales. First, the  initial mass in the
envelope which will  eventually fall inwards is reduced to 87\,\% of the total
required mass to form the star and excavate the bipolar outflow. The other
13\,\% is presumed to initially lie within a flattened disk. This yields the
values in the column 'infall mass' in Table\,\ref{parameters}.

In Fig.\,\ref{classzero}, we plot the sample summarised in
Table\,\ref{lbolvslhtwo} for Class\,0 sources for which a correlation of the
source properties with the outflow luminosity in the 1\,--\,0\,S(1) line of
H$_2$ has been investigated. Note that this sample includes warmer and less
luminous protostars than in the ISOPHOT sample investigated above. According to
the tracks, this corresponds to a wide range in final stellar masses. The
lowest mass star forming here is found to be  VLA\,1623 (assuming
T$_{bol}$\,=\,35\,K) which will reach just 0.07\,M$_\odot$, owing to its low
bolometric luminosity of only 1\,L$_\odot$ (Andr\'e et al. \cite{awb00}). The
low final mass is a result of this version of the evolutionary scheme employed,
for which we maintain the same accretion timescale  but alter the accretion
rate to generate the tracks. This implies that the final mass is nearly
proportional to the peak accretion luminosity. Future statistical studies will
lead to revisions of this first model.

The simplest form of the unifying model, assumed here, is that a fraction of
the jet power is instantaneously dissipated in shock waves, while the bipolar
outflow is a time-averaged recording of the momentum outflow. To model the
outflow, we have previously employed the H$_2$ luminosity, L(H$_2$), which we 
estimate to be ten times the 1\,--\,0\,S(1) luminosity. This is consistent with
expectations from shock physics and allows a comparison with previous diagrams
presented by Stanke \cite{stanke00} and Smith \cite{s02}. Here, however, we
shall use the jet power itself as the comparison parameter. For the comparison,
we assume that the observed emission is produced in the warm shocks where the
jets dissipate their energy, L$_{shock}$. 

The fraction of the jet power dissipated in molecular hydrogen lines is taken
to be 2\,\%. This is consistent with numerical simulations and bow shock
modelling which predict, typically, 10\,\% of the infrared radiation from
shocks in dense clouds to be in the form of H$_2$ lines. We also assume that
80\,\% of the jet energy is hidden by just under two magnitudes of K-band
extinction. The shock power has also been estimated from the far-infrared lines
of CO, OI, OH, and H$_2$O, measured by ISO (Giannini et al. \cite{gnl01}).
Here, we shall assume that these lines in total, within the ISO-LWS beam, also
represent 2\,\% of the jet power, L$_{shock}$. We thus increase estimated H$_2$
and sub-mm luminosities by 50 to yield the displayed values. While these
approximations are far from ideal, they permit us to determine if the
evolutionary scheme is plausible.

Figure\,\ref{h2-lbol0} demonstrates that the Class\,0 protostars possess almost
exclusively high ratios of L($_{shock}$)/L$_{bol}$. Two objects, however,
RNO\,15\,FIR and Ser-FIRS1, lie well within the model Class\,1 regime.
Environmental factors could cause a downward shift of the data points,
including higher extinction or radiative shocks which are less efficient in
H$_2$ vibrational excitation. Furthermore, these data are consistent with the
same model tracks fitted to the bolometric luminosity--temperature data. The
lack of a correlation in the data is thus put down to the combination of the
distributions in both mass and age.

 %  
 %  One of the two sources well within the Class\,1 regime is VLA\,1623 which may
 %  be obscured by up to 5 magnitudes of K-band extinction (Davis \& Eisl\"offel
 %  \cite{de95}). This would lead to a 16 times higher outflow luminosity, placing
 %  VLA\,1623 right into the Class\,0 regime. 
 %  

\begin{table}
\caption{\label{parameters}
Parameters derived for the seven objects from the model evolutions.  The
minimum mass is the total mass with density distributed as $\rho \propto
r^{-3/2}$ necessary to provide an optically thick sphere out to a radius
R$_{bol}$, corresponding to the observed T$_{bol}$. The infall mass is the
envelope mass which remains to be accreted (a part of which will escape in the
jets), and the envelope mass is the total mass predicted on projecting the
distribution out to a radius corresponding to the chosen  ambient temperature
of 24\,K. The model mass accretion rate decreases as t$^{-7/4}$ on a 30,000\,yr
timescale. The age is given in 10$^3$ years, the masses are in solar masses.
For comparison we give the measured values for the envelope masses from the
literature in column M$_{env}$. The values correspond to the M$_{env}$ column
in Table\,\ref{lbolvslhtwo}. The envelope mass for L\,1211\,FIRS2 is taken from
Tafalla et al. (1999).
}
\begin{center}
\renewcommand{\tabcolsep}{3pt} 
\begin{tabular}{lcccccrr}
Object & Age & Mass & Final & Min. & Infall & Env. & M$_{env}$ \\
 &  &  & mass & mass & mass & mass &  \\
\noalign{\smallskip}
\hline
\noalign{\smallskip}
RNO\,15\,FIR  &29.7 &0.10 &0.5 &0.05 &0.32 & 1.0 & 0.9 \\
HH\,211-MM    &24.0 &0.06 &0.3 &0.19 &0.26 & 1.4 & 1.5 \\
L\,1157       &27.0 &0.10 &0.5 &0.10 &0.35 & 1.3 & 0.5 \\
IC\,1396\,W   &24.9 &0.21 &1.2 &0.50 &0.88 & 4.2 & --  \\
L\,1211       &24.0 &0.43 &2.6 &1.53 &1.91 &10.3 & 0.8 \\
L\,1211\,FIRS2&22.5 &0.21 &1.4 &1.42 &1.06 & 6.9 & 2.1 \\
Cep\,E        &30.6 &0.97 &4.2 &0.37 &2.83 & 8.1 & 7.0 \\
\noalign{\smallskip}
\hline
\noalign{\smallskip}
\end{tabular}
\end{center}
\end{table}

In addition, previously measured Class\,1 outflows almost all lie below the
predicted Class border line, as shown in Fig.\,\ref{h2-lbol1}. Note, however,
that for many of the H$_2$ flows in Orion detected by Stanke \cite{stanke00}
only upper limits for the bolometric luminosity are available. Nevertheless,
the division of the two Classes with the model straight line where the
protostar has acquired half the final stellar mass, is evident. A similar
difference in outflow luminosity between Class\,0 and Class\,1 sources has been
found by Bontemps et al. \cite{batc96}.

\begin{figure}
\includegraphics[width=8.5cm, bb=0 9 516 428]{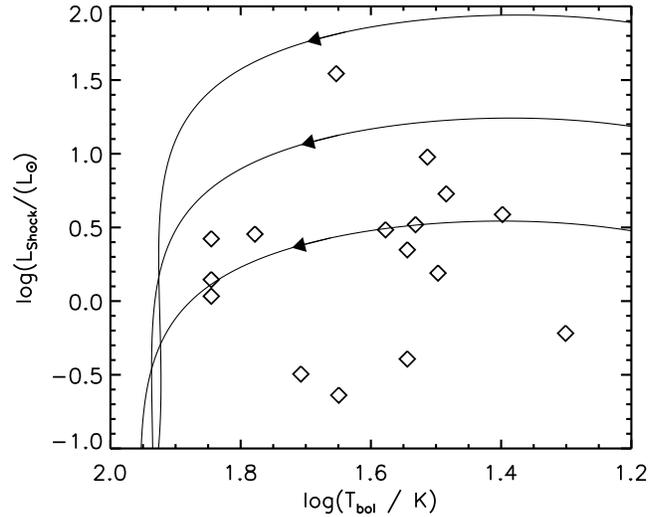}
\caption{\label{lshock-tbol} The derived outflow shock luminosity versus the
bolometric source temperature for the Class\,0 sources listed in
Table\,\ref{lbolvslhtwo}. The model tracks are for the same three models
presented in Fig\,\ref{classzero}.}
\end{figure}

Figures\,\ref{lshock-tbol} demonstrates that the model is also consistent with
the envelope properties. The main exceptions apparent from this diagram are a
group of low luminosity H$_2$  objects. This suggests that the extinction for
these sources may far exceed the fiducial two magnitudes.

%%%%%%%%%%%%%%%%%%%%%%%%%%%%%%%%%%%%%%%%%%%%%%%%%%%%%%%%%%%%%%%%%%%%%%%%%%%%%%
%%%%%%%%%%%%%%%%%%%%%%   Conclusions    %%%%%%%%%%%%%%%%%%%%%%%%%%%%%%%%%%%%%%
%%%%%%%%%%%%%%%%%%%%%%%%%%%%%%%%%%%%%%%%%%%%%%%%%%%%%%%%%%%%%%%%%%%%%%%%%%%%%%

\section{Conclusions}
 
We have observed the spectral energy distributions for the six deeply embedded
objects Cep\,E, L\,1211, IC\,1396\,W, L\,1157, HH\,211-MM, and RNO\,15\,FIR in
the far-infrared with ISO. The inferred temperatures and L$_{smm}$/L$_{bol}$
ratios confirm the Class\,0 nature of four of these sources, within the errors.
Employing an evolutionary scheme, we are able to estimate the age, surrounding
mass and the current and final mass of these sources. These estimates are,
however, model dependent. Two sources,  Cep\,E and L\,1211 appear to develop
into intermediate mass stars, while the others will become solar mass stars or
lower mass objects. The comparison of the ISOPHOT and LWS observations for
three of the sources reveals the existense of emission from cold dust in the
immediate vicinity of the objects.

A comparison of the luminosity in the 1\,--\,0\,S(1) line of H$_2$ of the
related outflows for 16 Class\,0 sources, with the source bolometric
luminosity, bolometric temperature, and envelope mass was done. We found no
statistically significant correlation of the outflow luminosity with each of
these source parameters. This could be due to the H$_2$ luminosity mainly
depending on the local properties of the surrounding gas.

The unifying scheme, however, explains the lack of correlations as due to
evolutionary effects. Furthermore, the scheme which involves a redistribution
of mass between envelope, disk, protostar, jets and outflow, accounts for the
differences in source properties according to the Class.

\section*{Acknowledgement}

We thank Manfred Stickel from MPIA for his help with the data reduction of the
PHOT data. We thank Alex Rosen for a critical reading of the manuscript and the
Department of Culture, Arts and Leisure, Northern Ireland for financial
support. Jochen Eisl\"offel and Dirk Froebrich received financial support from
the DLR through Verbundforschung grant 50\,OR\,9904\,9. \\ The ISOPHOT data
presented in this paper were reduced using PIA, which is a joint development by
the ESA Astrophysics Division and the ISOPHOT Consortium with the collaboration
of the Infrared Processing and Analysis Center (IPAC). Contributing ISOPHOT
Consortium institutes are DIAS, RAL, AIP, MPIK and MPIA.\\ The ISO Spectral
Analysis Package (ISAP) is a joint development by the LWS and SWS Instrument
Teams and Data Centers. Contributing institutes are CESR, IAS, IPAC, MPE, RAL
and SRON. 

%\appendix
%==============================================================================
\section*{Appendix}

We present and test a model based on the transfer of gas between components. We
take a spherical envelope of gas and dust,  and prescribe an accretion rate
from the inner edge of the envelope onto a disk. Note that we assume a
centrifugal barrier at 30\,AU, which defines the inner envelope -- outer disc
transition.  The accretion disk processes most of the mass onto the protostar
and a fraction into twin jets. The speed of the jets is assumed to be a fixed
fraction of the escape speed from the protostellar surface. 

The accretion rate from the envelope is taken to
increase exponentially for a short period before decreasing as a power
law through  the Class\,0, 1 and 2 phases. The zero point of time is thus
defined as the moment when accretion starts and, simultaneously, a central
hydrostatic object forms. The accretion rate is
\begin{equation}
   \dot  M_a(t) = 
   \dot  M_o (e/{\alpha})^{\alpha}(t/t_o)^{-\alpha} \exp(-t_o/t).
\label{eqnpower}
\end{equation}
Energy release through accretion and contraction are
included.  In the models shown, the peak accretion rate is reached at
t$_o/\alpha = 17,000$\,yr, and the power law index is $\alpha = 7/4$, on a
t$_o$ = 30,000\,yr timescale. The accreted mass is predominantly accrued by the
growing protostars. The fraction $\epsilon(t)$ which escapes through twin jets
reaches a maximum of $\eta = 0.2$ at the peak accretion time:
\begin{equation}
     \epsilon = \eta \left[\frac{\dot M_a(t)}{\dot M_o}\right]^\zeta
\label{eqn-hm}
\end{equation}
where $\zeta = 2$ is found to be appropriate. Hence the mass left over, which 
accretes onto the core to form the star is
\begin{equation}
M_*(t) = \int_0^t  (1\,-\,\epsilon)\,\dot M_a.
\end{equation}

To form a star like the Sun, this model will  provide an early accretion peak
in which $\dot M_a \sim 10^{-4}M_{\odot}\,{\rm yr}^{-1}$ for 10$^4$ years, and
eventually  becoming $\dot M_a \sim 10^{-7}M_{\odot}\,{\rm yr}^{-1}$ for
10$^6$\,years, corresponding to Class\,0 and Class\,2 or Classical T\,Tauri
stars, respectively. The power-law has substantial observational support
(Calvet et al. \cite{chs00}).

We previously modelled the envelope evolution by assuming mass conservation. 
Here, we make two significant adjustments in order to model the new data.
First, the  initial mass in the envelope which will  eventually fall inwards is reduced to
87\% of the total required mass to form the star and excavate the bipolar
outflow. The other 13\,\% is presumed to initially lie within a flattened disk.
This yields the values in the column 'infall mass' in Table\,\ref{parameters}.
The total mass can be written analytically in terms of an incomplete Gamma
function on integrating Eqn.\,\ref{eqnpower}:
\begin{equation}
   M_{infall}(t) = 
   \dot  M_ot_o (e/{\alpha})^{\alpha}\left[1\,-\,{\Gamma}
                     (\alpha-1,t_o/t)\right].
\end{equation}
Secondly, we find that the low bolometric temperatures of Class\,0 protostars 
can only be attained by introducing an additional mass component to the
envelope. In confirmation of the results of Myers et al. \cite{macs98},  we
find that this extra mass is lost on a shorter timescale than the protostellar
accretion timescale. The bolometric temperature is calculated according to the
optically thick case of Myers et al. \cite{macs98}. We thus extend the
opacity law approximation employed up to 60 to 120\,$\mu$m with the same form
and take the optically thick envelope throughout the early evolutionary stages.
We have thus found here that an envelope mass
\begin{equation}
   M_{env}(t) = 
    M_{inf} \cdot (0.87 + \mu (t/t_o)^{-2 \alpha})
\label{envel}
\end{equation}
where $\mu = 2$ provides bolometric temperatures, timescales and masses
consistent with the observed samples. 

The envelope mass provides a testable prediction. This mass is not strongly
dependent on the evolutionary path but is necessary to provide the optical
depth out to a sufficiently large radius to permit the measured low bolometric
temperature. The total mass is dominated by the outer regions of the envelope,
while the total optical depth is controlled by the inner region (for all
plausible density distributions such as $\rho \propto r^{-3/2}$, as assumed
here). Hence, the mass is sensitive to the extent of the envelope. For this
reason, we present three determinations of the envelope mass in
Table\,\ref{parameters}. Masses derived from submillimetre observations  yield
quite  low extended masses (Andr{\'e} et al. \cite{awb00}),  consistent with
the absence of more mass than necessary to form the star and feed the jets
(Smith \cite{s00}). It is clear that both the observationally derived mass and
model mass are sensitive to chosen physical parameters and both will need
refining.

%==============================================================================

\label{lastpage}


\begin{thebibliography}{}

%%%%%%%%%%%%%%%%%%%%%%%%%%%%%%%%%%%%%%%%%%%%%%%%%%%%%%%%%%%%%%%%%%%%%%%%%%%%%%
%%%%%%%%%%%%%%%%%%%%%%      Papers      %%%%%%%%%%%%%%%%%%%%%%%%%%%%%%%%%%%%%%
%%%%%%%%%%%%%%%%%%%%%%%%%%%%%%%%%%%%%%%%%%%%%%%%%%%%%%%%%%%%%%%%%%%%%%%%%%%%%%

\bibitem[2000]{awb00}
Andr{\'e}, P., Ward-Thompson, D., Barsony, M. 2000, in Protostars and Planets
IV, 59 
%From prestellar cores to protostars: The initial conditions of star formation

\bibitem[2002]{ar02}
Anglada, G., Rodr\'{\i}guez, L.F., 2002, Revista Mexicana de Astronom\'{\i}a y
Astrof\'{\i}sica, 38, 12
%VLA detection of the exciting sources of the molecular outflows associated 
%with L1448 IRS2, IRAS 05327+3404, L43, IRAS 22142+5206, L1211 and IRAS
%23545+6508 

\bibitem[1998]{bwao98}
Barsony, M., Ward-Thompson, D., Andr\'e, P., O\'{}Linger, J., 1998, AJ, 509, 
733
%Protostars in Perseus: outflow-induced fragmentation

\bibitem[1996]{batc96} 
Bontemps, S., Andr\'e, P., Terebey, S., Cabrit, S., 1996, A\&A, 311, 858
%Evolution of outflow activity around low-mass embedded young stellar objects

\bibitem[2000]{chs00} 
Calvet, N., Hartmann, L.W., Strom, S.E., 2000, in Protostars \& Planets IV, ed.
V. Mannings et al., (Tucson, U. of Arizona Press), 377
%Evolution of Disk Accretion

\bibitem[2001]{cwknrs01}
Chini, R., Ward-Thompson, D., Kirk, J.M., Nielbock, M., Reipurth, B., Sievers,
A., 2001, A\&A, 369, 155
%MM/Submm images of Herbig-Haro energy sources and candidate protostars

\bibitem[1996]{caa_etal96}
Clegg, P.E., Ade, P.A.R., Armand, C., et al., 1996, A\&A, 315, L38
%The ISO Long-Wavelenth Spectrometer.

\bibitem[1995]{de95}
Davis, C.J., Eisl\"offel, J., 1995, A\&A, 300, 851
%Near-infrared imaging in H2 of molecular (CO) outflows from young stars

\bibitem[1998]{dsm98} 
Davis C.J., Smith M.D., Moriarty-Schieven G.H., 1998, MNRAS, 299, 825
% first mention of unification scheme

\bibitem[1997]{drec97}
Davis, C.J., Ray, T.P., Eisl\"offel, J., Corcoran, D., 1997, A\&A, 324, 263
%Near-IR imaging of the molecular outflows in HH24-26, L1634(HH240-241)
%L1660(HH72) and RNO15FIR.

\bibitem[2000]{e00}
Eisl\"offel, J., A\&A, 2000, 354, 236
%Parsec-scale molecular H2 outflows from young stars

\bibitem[2004]{ef04}
Eisl\"offel, J., Froebrich, D., 2004, in prep.
%outflows in the serpens cloud....

\bibitem[2003]{efsm03}
Eisl\"offel, J., Froebrich, D., Stanke, T., McCaughrean, M.J., 2003, ApJ, 
in press
%Molecular outflows in the young open cluster IC 348

\bibitem[1996]{esdr96}
Eisl\"offel, J., Smith, M.D., Davis, C.J., Ray, T.P., 1996, AJ, 112, 2086
%Molecular Hydrogen in the Outflow from CepE

\bibitem[2004]{ezf04}
Eisl\"offel, J., Ziener, R., Froebrich, D., 2004, in prep.
%search for outflows in OrionB...

\bibitem[2004]{fe04}
Froebrich, D., Eisl\"offel, J., 2004, in prep.
%outflow properties of the L???? objects

\bibitem[2003]{fs03}
Froebrich, D., Scholz, A., 2003, A\&A, in press
%Young stars and outflows in the globule IC1396W

\bibitem[2002]{fse02}
Froebrich, D., Smith, M.D., Eisl\"offel, J., 2002, A\&A, 385, 239
%Far-infrared spectroscopy across the asymmetric bipolar outflows from
%CepheusA and L1448

\bibitem[2001]{fze01}
Froebrich, D., Ziener, R., Eisl\"offel, J., AG Abstr. Ser., 18, 25
%An unbiased search for molecular hydrogen outflows in the OrionB star 
%forming region

\bibitem[1997]{g_etal97}
Gabriel, C., et al., 1997, in ASP Conf. Ser. Vol. 125, Astronomical Data
Analysis Software and Systems (ADASS) VI, ed. G. Hunt \&  H.E. Payne, (San
Francisco ASP), 108
%The ISOPHOT Interactive Analysis PIA, a calibration and scientific analysis
%tool.

\bibitem[2001]{gnl01}
Giannini, T., Nisini, B., Lorenzetti, D., 2001, ApJ, 555, 40
%Far-Infrared Investigation of Class 0 Sources: Line Cooling

\bibitem[1999]{gg99}
Gueth, F., Guilloteau, S., 1999, A\&A, 343, 571
%The jet-driven molecular outflow of HH\,211

\bibitem[1997]{ggdb97}
Gueth, F., Guilloteau, S., Dutrey, A., Bachiller, R., 1997, A\&A, 323, 943
%Structure and kinematics of a protostar: mm-interferometry of L 1157

\bibitem[1997]{husm97}
Hirose, S., Uchida, Y., Shibata, K., Matsumoto, R., 1997, PASJ, 49, 193
%Disk Accretion onto a Magnetized Young Star and Associated Jet Formation

\bibitem[1996]{ksa_etal96}
Kessler, M.F., Steinz, J.A., Anderegg, M.E., et al., 1996, A\&A, 315, L27
%The infrared Space Observatory (ISO) mission.

\bibitem[1997]{lh97}
Ladd, E.F., Hodapp, K.-W., 1997, ApJ, 475, 749
%A double outflow from a deeply embedded source in cepheus

\bibitem[1999]{l99}
Laureijs, R.J., 1999, Point spread function fractions related to the ISOPHOT
C100 and C200 arrays, ISO-Data centre, Astrophysics Division, ESA, Villafranca
Spain, http://www.iso.vilspa.esa.es/users/expl\_lib/PHT/c200fpsf02.ps.gz
%Point spread function fractions related to the ISOPHOT C100 and C200 arrays

\bibitem[1996]{lel96}
Lefloch, B., Eisl\"offel, J., Lazareff, B., 1996, A\&A, 313, L17
%The remarkable Class 0 source CEP E

\bibitem[1996]{lka_etal96}
Lemke, D., Klaas, U., Abolins, J., et al., 1996, A\&A, 315, L64
%ISOPHOT-capabilities and performance.

\bibitem[1994]{mrz94}
McCaughrean, M.J., Rayner, J.T., Zinnecker, H., 1994, ApJ, 436, L189
%Discovery of a molecular Hydrogen jet near IC348

\bibitem[2001]{mnmtcs01}
Moro-Mart{\'\i}n, A., Noriega-Crespo, A., Molinari, S., Testi, L., Chernicharo,
J., Sargent, A. 2001, A\&A, 555, 146
%Infrared and Millimetric study of the young outflow Cepheus E

\bibitem[2001]{ma01}
Motte, F., Andr\'e, P., 2001, A\&A, 365, 440
%The circumstellar environment of low-mass protostars: A millimeter continuum
%mapping survey

\bibitem[1998]{macs98}
Myers, P.C., Adams, F.C., Chen, H., Schaff, E., 1998, ApJ, 492, 703
%The LBT luminosity-bolometric-temperature model

\bibitem[1997]{op97}
Ouyed, R., Pudritz, R.E., 1997, ApJ, 482, 712
%Numerical Simulations of Astrophysical Jets from Keplerian Disks. I.
%Stationary Models

\bibitem[2003]{reh03}
Rengel, M., Eisl\"offel, J., Hodapp, K.-W., 2003, in prep.
%Submillimetre SCUBA imaging of deeply embedded outflow sources and Class 0
%sources 

\bibitem[2001]{rfeh01}
Rengel, M., Froebrich, D., Eisl\"offel, J., Hodapp, K., 2001, AG Abstr. Ser.
18, 66
%Submillimetre Imaging of Deeply Embedded Outflow Sources and Class 0 Sources

\bibitem[2002]{rfhe02}
Rengel, M., Froebrich, D., Hodapp, K., Eisl\"offel, J., 2002, In: The Origins
of Stars and Planets: The VLT View, Jo{\~a}o Alves \& Mark McCaughrean (ed.)
%Far-Infrared and Submillimetre imaging of deeply embedded outflow sources

\bibitem[1996]{saraceno96}
Saraceno P., Andr\'e P., Ceccarelli C., Griffin M., Molinari S.,
1996, A\&A 309 827 
%An evolutionary diagram for young stellar objects.

\bibitem[1994]{snowrl94}
Shu, F., Najita, J., Ostriker, E., Wilkin, F., Ruden, S., Lizano, S., 1994,
ApJ, 429, 781
%Magnetocentrifugally driven flows from young stars and disks. 1: A generalized
%model

\bibitem[1998]{s98} 
Smith M.D., 1998, Ap\&SS, 261, 169
%The Evolution of Young Stars, Protostellar Jets & Bipolar Outflows - a
%Unification Scheme 

\bibitem[2000]{s00}
Smith, M.D., 2000, IrAJ, 27, 25
%Evolutionary schemes for protostars, proto brown dwarfs and their
%environments.

\bibitem[2002]{s02}
Smith, M.D., 2002, In: The Origins of Stars and Planets: The VLT View, Jo{\~a}o
Alves \& Mark McCaughrean (ed.)
%The Unification Scheme and some techniques for tracking the evolution of
%protostars

\bibitem[1990]{sb90}
Smith, M.D., Brand, P.W.J.L., 1990, MNRAS, 245, 108
%H2 profiles of C-type bow shocks

\bibitem[2003]{sfe03}
Smith, M.D., Froebrich, D., Eisl\"offel, J., 2003, ApJ, 592, 245
%Multiwavelength Spectroscopy of the Bipolar Outflow from Cepheus E

\bibitem[2000]{stanke00}
Stanke, T., 2000, Ph.D. thesis (AIP, Potsdam)
%an unbiased h2 servey for molecular outflows in OrionA

\bibitem[1974]{ssc74}
Strom, S.E., Strom, K.A., Carrasco, L., 1974, PASP, 86, 798
%A study of the young cluster IC 348
 
\bibitem[1999]{tmmb99}
Tafalla, M., Myers, P.C., Mardones, D., Bachiller, R., 1999, A\&A, 348, 479
%A cluster of young stellar objects in L1211

\bibitem[2000]{ybsbb00}
Yu, K.C., Billawala, Y., Smith, M.D., Bally, J., Butner, H., 2000, AJ, 120,
1974 

\bibitem[1998]{zmr98}
Zinnecker, H., McCaughrean, M.J., Rayner, J.T., 1998, Nature, 394, 862
%A symmetrically pulsed jet of gas from an invisible protostar in Orion

%%%%%%%%%%%%%%%%%%%%%%%%%%%%%%%%%%%%%%%%%%%%%%%%%%%%%%%%%%%%%%%%%%%%%%%%%%%%%%
%%%%%%%%%%%%%%%%%%%%%   Papers   %%%%%%%%%%%%%%%%%%%%%%%%%%%%%%%%%%%%%%%%%%%%%
%%%%%%%%%%%%%%%%%%%%%%%%%%%%%%%%%%%%%%%%%%%%%%%%%%%%%%%%%%%%%%%%%%%%%%%%%%%%%%

\end{thebibliography}
\end{document}